{}

\documentclass[11pt,a4paper]{article}

\newif\ifpublic\publictrue
\newif\iffancy\fancytrue

\usepackage[a4paper,text={160mm,247mm},centering]{geometry}
\usepackage{setspace}
\usepackage{rotating}
\usepackage[margin=10pt,font=small,labelfont=bf, labelsep=colon]{caption} [2022/02/20]

\usepackage{amsmath, amssymb, amsfonts, amsthm, geometry}

\newtheorem*{theorem*}{Theorem}

\usepackage{enumitem}

\usepackage[usenames,dvipsnames,table]{xcolor}
\definecolor{mygreen}{rgb}{0,0.4,0}
\definecolor{myblue}{rgb}{0,0.0,0.4}
\definecolor{refrcolor}{rgb}{0,0.4,0}
\definecolor{cgreen}{rgb}{0,0.7,0}
\definecolor{ecolor}{rgb}{.52,.03,.06}
\definecolor{bgcolor}{rgb}{.96,.95,.80}
\definecolor{bgcolordark}{rgb}{.80,.80,.67}
\definecolor{faint}{rgb}{.80,.80,.80}

\usepackage{tikz}
\usetikzlibrary{decorations.markings}
\usetikzlibrary{decorations.pathmorphing}
\usepackage{graphicx}
\usepackage{subcaption}
\usepackage{caption}
\usepackage[theorems]{tcolorbox}
\newtcolorbox{mymathbox}[1][]{colframe=black,colback=white, sharp corners, #1}
\usepackage{pdfpages}
\usepackage{bm}
\tikzset{snake it/.style={decorate, decoration=snake}}

\makeatletter
\providecommand*{\shuffle}{%
  \mathbin{\mathpalette\shuffle@{}}%
}
\newcommand*{\shuffle@}[2]{%
  \sbox0{$#1\vcenter{}$}%
  \kern .15\ht0 
  \rlap{\vrule height .25\ht0 depth 0pt width 2.5\ht0}%
  \raise.1\ht0\hbox to 2.5\ht0{%
    \vrule height 1.75\ht0 depth -.1\ht0 width .17\ht0 %
    \hfill
    \vrule height 1.75\ht0 depth -.1\ht0 width .17\ht0 %
    \hfill
    \vrule height 1.75\ht0 depth -.1\ht0 width .17\ht0 %
  }%
  \kern .15\ht0 
}
\makeatother

\makeatletter
\g@addto@macro\bfseries{\boldmath}
\makeatother

\usepackage{xparse}

\ExplSyntaxOn
\NewDocumentCommand{\Gtargz}{m m}
{
 \Gt\left(\begin{smallmatrix}
 \Gtargz_print:n {#1} \\
 \Gtargz_print:n {#2}
 \end{smallmatrix};z\right)
}
\seq_new:N \l_Gtargz_list_seq
\cs_new_protected:Npn \Gtargz_print:n #1
{
  \seq_set_split:Nnn \l_Gtargz_list_seq { , } { #1 }
  \seq_use:Nn \l_Gtargz_list_seq { , & }
}
\ExplSyntaxOff

\ExplSyntaxOn
\NewDocumentCommand{\Gtargzt}{m m}
{
 \Gt\left(\begin{smallmatrix}
 \Gtargzt_print:n {#1} \\
 \Gtargzt_print:n {#2}
 \end{smallmatrix};z,\tau\right)
}
\seq_new:N \l_Gtargzt_list_seq
\cs_new_protected:Npn \Gtargzt_print:n #1
{
  \seq_set_split:Nnn \l_Gtargzt_list_seq { , } { #1 }
  \seq_use:Nn \l_Gtargzt_list_seq { , & }
}
\ExplSyntaxOff
\newcommand{\SI}[1]{\Sel[#1]}

\ExplSyntaxOn
\NewDocumentCommand{\SIE}{m m}
{
\SelE\!\Big[\begin{smallmatrix}
 \SI_print:n {#1} \\
 \SI_print:n {#2}
 \end{smallmatrix}\Big]
}
\seq_new:N \l_SI_list_seq
\cs_new_protected:Npn \SI_print:n #1
{
  \seq_set_split:Nnn \l_SI_list_seq { , } { #1 }
  \seq_use:Nn \l_SI_list_seq { , & }
}
\ExplSyntaxOff

\usepackage[pdfencoding=auto,bookmarks=true,hyperfigures=true]{hyperref}%
\PassOptionsToPackage{unicode}{hyperref}
\usepackage{graphbox}
\usepackage{float}
\usepackage[nosort]{cite}
\usepackage{lmodern}
\usepackage{amsbsy}
\usepackage[utf8]{inputenc}
\usepackage[font=small,labelfont=bf]{caption}

\numberwithin{equation}{section}

\newcommand{\eqn}[1]{eq.~\eqref{#1}}
\newcommand{\Eqn}[1]{Equation~\eqref{#1}}
\newcommand{\eqns}[2]{eqs.~\eqref{#1} and~\eqref{#2}}

\newcommand{\rcite}[1]{ref.~\cite{#1}}
\newcommand{\rcites}[1]{refs.~\cite{#1}}

\providecommand{\href}[2]{#2}

\makeatletter
\def\mr@ignsp#1 {\ifx\:#1\@empty\else #1\expandafter\mr@ignsp\fi}%
\newcommand{\multiref}[1]{\begingroup
\xdef\mr@no@sparg{\expandafter\mr@ignsp#1 \: }%
\def\mr@comma{}%
\@for\mr@refs:=\mr@no@sparg\do{\mr@comma\def\mr@comma{,}\ref{\mr@refs}}%
\endgroup}
\renewcommand{\eqref}[1]{(\multiref{#1})}
\makeatother

\makeatletter
\newcommand{\namedref}[2]{#1~\hyperref[#2]{\ref*{#2}}}
\newcommand{\secref}{\@ifstar{\namedref{Section}}{\namedref{section}}}
\newcommand{\subsecref}{\@ifstar{\namedref{Subsection}}{\namedref{subsection}}}
\newcommand{\appref}{\@ifstar{\namedref{Appendix}}{\namedref{appendix}}}
\newcommand{\tabref}{\@ifstar{\namedref{Table}}{\namedref{table}}}
\newcommand{\figref}{\@ifstar{\namedref{Figure}}{\namedref{figure}}}
\newcommand{\Figref}{\@ifstar{\namedref{Figure}}{\namedref{Figure}}} 
\makeatother

\providecommand{\hypersetup}[1]{}
\providecommand{\texorpdfstring}[2]{#1}

\hypersetup{plainpages=false}
\hypersetup{pdfpagemode=UseNone}
\hypersetup{bookmarksnumbered=true}
\hypersetup{pdfstartview=FitH}
\hypersetup{colorlinks=false}
\hypersetup{citebordercolor={0.7 0.7 1}}
\hypersetup{urlbordercolor={.4 .8 1}}
\hypersetup{linkbordercolor={1 .8 .6}}
\hypersetup{colorlinks=true, urlcolor=[rgb]{0.13,0.30,0.45}, linkcolor=[rgb]{0.13,0.30,0.45}, citecolor=[rgb]{0.55,0.0,0.05}}
\makeatletter
\let\@keywords\@empty
\let\@subject\@empty
\providecommand{\keywords}[1]{\gdef\@keywords{#1}}
\providecommand{\subject}[1]{\gdef\@subject{#1}}
\def\thetitle{\@title}
\def\theauthor{\@author}
\def\thesubject{\@subject}
\def\thedate{\@date}
\def\thekeywords{\@keywords}
\makeatother
\AtBeginDocument{
\hypersetup{pdftitle={\thetitle}}%
\hypersetup{pdfauthor={\theauthor}}%
\hypersetup{pdfsubject={\thesubject}}%
\hypersetup{pdfkeywords={\thekeywords}}%
}

\newif\ifnote 
\notetrue

\allowdisplaybreaks[1]

\let\qed\relax\newcommand{\qed}
{\hfill\ensuremath{\Box}}

\newcommand{\pd}{\partial}

\newcommand{\ve}{\varepsilon}

\newcommand{\SL}{\mathrm{SL}}

\newcommand{\mto}{\rightarrow}
\newcommand{\te}{\textrm}

\newcommand{\ap}{\alpha'}

\newcommand{\zb}{\bar{z}}

\newcommand{\ZQ}{\mathbb Q}
\newcommand{\ZR}{\mathbb R}

\newcommand{\CA}{\mathcal{A}}

\newcommand{\CF}{\mathcal{F}}       
\newcommand{\CG}{\mathcal{G}} 

\newcommand{\CI}{\mathcal{I}}

\newcommand{\CL}{\mathcal{L}}       
\newcommand{\CN}{\mathcal{N}}      
\newcommand{\CM}{\mathcal{M}}

\DeclareMathOperator{\Gammafn}{\Gamma}

\DeclareMathOperator{\Gt}{\tilde{\Gamma}}

\DeclareMathOperator{\Sel}{S}
\DeclareMathOperator{\SelE}{S^E}

\DeclareMathOperator{\zm}{\zeta}

\usepackage{nicefrac}

\newcommand{\Nfour}{${\cal N}{=}\,4$ {}}

\newcommand{\zpih}{z{+}\tfrac{i}{2}}

\newcommand{\xp}{\chi_+}
\newcommand{\xm}{\chi_-}

\vspace*{0.2cm}
\title{\textbf{
    A KLT-like construction for multi-Regge amplitudes
    }}
    \author{Konstantin Baune\texorpdfstring{$^{\,\textit{a}}$}{},
	    Johannes Broedel\texorpdfstring{$^{\,\textit{a}}$}{}
}
\date{\today}

\begin{document}

\pdfbookmark[1]{Title Page}{title} 
\thispagestyle{empty}
\vspace*{2.4cm}
\begin{center}%
  \begingroup\LARGE\bfseries\thetitle\par\endgroup
\vspace{1.0cm}

\begingroup\large\theauthor\par\endgroup
\vspace{9mm}
\begingroup\itshape
$^{\te{a}}$Institute for Theoretical Physics, ETH Zurich\\Wolfgang-Pauli-Str.~27, 8093 Zurich, Switzerland
\par\endgroup
\vspace*{7mm}

\begingroup\ttfamily
baunek@ethz.ch, jbroedel@ethz.ch
\par\endgroup

\vspace*{1.5cm}

\textbf{Abstract}\vspace{5mm}

\begin{minipage}{13.4cm}
Inspired by the calculational steps originally performed by Kawai, Lewellen and Tye, we decompose scattering amplitudes with single-valued coefficients obtained in the multi-Regge-limit of \Nfour super-Yang--Mills theory into products of scattering amplitudes with multi-valued coefficients. We consider the simplest non-trivial situation: the six-point remainder function complementing the Bern--Dixon--Smirnov ansatz for multi-loop amplitudes. 

Utilizing inverse Mellin transformations, all single-valued amplitude components can indeed be decomposed into multi-valued amplitude components. Although the final expression is very similar in structure to the Kawai--Lewellen--Tye construction, moving away from the highly symmetric string scenario comes with several imponderabilities, some of which become more pronounced when considering more than six external legs in the remainder function. 
\end{minipage}
\vspace*{4cm}
\end{center}

\newpage
\setcounter{tocdepth}{2}
\tableofcontents

\newpage

\section{Introduction}
\label{sec:introduction}

Modern amplitude calculations rely on two main structures for easing the process: \textit{recursion relations}, which allow to infer high-multiplicity many-loop amplitudes from those with fewer legs and fewer loops, and \textit{doubling relations} which allow to ``square'' scattering amplitudes of one theory (usually a gauge theory) in order to obtain corresponding objects in another theory (a gravitational theory in most cases). In conjunction with the use of polylogarithms and their single-valued analogues, these insights and associated algorithms have triggered enormous progress in calculating scattering amplitudes in various gauge and string theories during the last decades. 

Especially the collection of gluon scattering amplitudes in planar \Nfour super-Yang--Mills (sYM) theory in multi-Regge kinematics (MRK) \cite{Lipatov:1993yb, Lipatov:1994xy, Faddeev:1994zg, DelDuca:2013lma} exhibit recursive relations: initiated by expressing higher-loop gluon amplitudes in planar \Nfour sYM theory in terms of the one-loop amplitude \cite{Anastasiou:2003}, a conjectured all-loop ansatz for an arbitrary number of external gluons was put forward by Bern, Dixon and Smirnov (BDS ansatz) in \rcite{Bern:2005iz}. While this ansatz is exact for four- and five-point amplitudes \cite{Drummond:2010}, a correction factor, called the \textit{remainder function}, needs to be included for more than five gluons \cite{two-loop,BFKLPomeron,reggecut,BFKLapproach}. The analytical form of the remainder function is restricted by dual conformal symmetry \cite{Drummond:2011} and it exhibits further structures that allow for iterative calculations (e.g.~\cite{BroedelSprenger}). The current final piece in this direction is the conjectured all-loop formula of the remainder function for an arbitrary number of external gluon legs \cite{DelDuca:2019tur}. A recent review of the current status of MRK gluon scattering amplitudes in planar $\mathcal{N}{=}\,4$ sYM theory can be found in ref.~\cite{DelDucaDixon}. 

While the recursion relations for gluon scattering amplitudes in MRK have already been exploited in many situations, we will discuss in this article, in what way the corresponding remainder functions in planar \Nfour sYM theory can be obtained from double-copy relations. This method appears reasonable, as those amplitudes (after dividing out the BDS ansatz) evaluate to single-valued polylogarithms exclusively \cite{Dixon:2012yy}, which in turn is a property usually attributed to gravitational theories. While \Nfour sYM theory is clearly not a gravitational theory, it will be discussed in what sense MRK paves the way towards the double-copy representation. 

Double-copy relations are usually explored starting from a set of open-string (or gauge-theory) amplitudes and comparing it to a set of available closed-string (or gravitational) amplitudes. In this article, we are going to follow the line of deduction of Kawai, Lewellen and Tye (KLT) \cite{KLT} and their predecessors~\cite{Dotsenko:1984nm,Dotsenko:1984ad}, who started from the closed-string integral and broke it apart into combinatorial sums over products of open-string amplitudes. Accordingly, we are going to start from the single-valued remainder function of multi-Regge amplitudes and rewrite it into a suitable ``square'' of multi-valued amplitudes.

The new multi-valued amplitudes turn out to be inverse Mellin transforms of different polygamma functions. We have not been able to find a theory producing those amplitudes right away, but recent progress in constructing amplitudes purely from their analytic properties \cite{Remmen:2021zmc,Duhr:2023his,Cheung:2023adk} nourish the hope to identify amplitude-like properties in the objects we found.

\medskip

This article is structured as follows: in \subsecref{sec:polylogszeta}, we briefly touch upon some basic concepts like polylogarithms and single-valuedness before presenting KLT's original approach to the double copy in \subsecref{sec:klt}. In \secref{sec:BFKLMRK} we are going to review the standard approach to calculating gluon amplitudes in \Nfour sYM theory in MRK. In particular the BDS ansatz for gluon scattering amplitudes in MRK will be revisited together with the six-point remainder function.

In section~\ref{sec:doublecopy} the analytic expression for the six-gluon remainder function in \Nfour sYM theory will be analyzed more carefully leading to the decomposition into two equivalent contributions which qualifies as a double-copy relation. Finally, after establishing this relation at the level of the six-point remainder function, we will provide an outlook in section~\ref{sec:7point} onto identifying similar relations for higher-point remainder functions by recycling insights from the six-point case. We are going to conclude and summarize in section~\ref{sec:results}. 

In three appendices we provide additional information: \appref{app:calculations} shows explicit calculations using our double-copy construction which reproduce known results. In \appref{app:proof} we provide several proofs used in \secref{sec:doublecopy}. Finally, in \appref{app:conSVHPL} we briefly review the construction of single-valued polylogarithms from \rcite{Brown:2004}. 


\section{Mathematical tools and a brief KLT review}

\subsection{Polylogarithms and multiple zeta values}
\label{sec:polylogszeta}
\paragraph{Polylogarithms.} As for many other calculations in \Nfour sYM theory, multiple polylogarithms and their single-valued analogues will be the main class of functions to be considered in the multi-Regge limit. Let us define the multiple polylogarithms
\begin{equation}
	\label{eqn:polylogdef}
	G(a_1,a_2,\ldots,a_n;z)=\int_0^z \frac{dt}{t-a_1}G(a_2,\ldots,a_n;t)\,,
\end{equation}
with $G(;z)=1$. Defining furthermore $G(0;z)=\log z$ implements shuffle regularization \cite{Deligne,Brown:ICM14}, where we follow the conventions of \rcite{Broedel:2019gba} in the following. While in general $a_i$ can be rational functions \cite{Panzer:2015ida} of the complex argument $z$, for our purposes it is sufficient to consider $a_i\in\{0,1\}$ exclusively. Corresponding to the two possible values of $a_i$, we define \textit{letters} $x_0$ and $x_1$ allowing to denote the polylogarithms from \eqn{eqn:polylogdef} by words in the subscript
\begin{equation}
	G_w(z)\text{ with }w\in X^\times\, ,
\end{equation}
where $X^\times$ is the set of all words built from the \textit{alphabet} $X\,{=}\,\{x_0,x_1\}$, which also contains the empty word $e$. Translation between the notation in \eqn{eqn:polylogdef} and the notation in terms of words is canonical, for example
\begin{equation}
	G(0,1,1;z)=G_{x_0x_1x_1}(z)\,.
\end{equation}
Employing the composition-of-paths formula, one can prove that polylogarithms $G$ satisfy the shuffle relations
\begin{equation}
	\label{eqn:shuffle1}
	G_{w_1}(z)G_{w_2}(z)=\sum_{w\in w_1\shuffle w_2} G_w(z)\,,
\end{equation}
for $w_1,w_2\in X^\times$ and $\shuffle$ the shuffle product\footnote{For letters $x_i,x_j\in X$ and words $e,u,v\in X^\times$ the shuffle product $\shuffle$ is recursively defined by $e\shuffle u=u\shuffle e=u$ and $(x_i u)\shuffle(x_j v)=x_i(u\shuffle(x_j v))+x_j((x_i u)\shuffle v)$.}.
Using definition \eqref{eqn:polylogdef}, one finds 
\begin{equation}
	\frac{\pd}{\pd z}G_{x_0w}(z)=\frac{G_w(z)}{z}\,,\quad \frac{\pd}{\pd z}G_{x_1w}(z)=\frac{G_w(z)}{z-1}\,.
\end{equation}
The class of polylogarithms defined in \eqn{eqn:polylogdef} together with the shuffle regularization prescription is linked to the usual polylogarithms via
\begin{equation}
	\label{eqn:polylogrelations}
	\log^n z = G(\underbrace{0,\ldots,0}_n;z)\,,\quad \operatorname{Li}_n(z)=-G(\underbrace{0,\ldots,0}_{n-1},1;z)\,,\quad H_w(z)=(-1)^{\#(x_1,w)}G_w(z)\,,
\end{equation}
where $H_w(z)$, $w\in X^\times$ are the harmonic polylogarithms discussed in \rcite{Remiddi:2000} and $\#(x_1,w)$ denotes the number of letters $x_1$ in $w$.

\paragraph{Multiple zeta values.} Evaluating multiple polylogarithms in \eqn{eqn:polylogdef} at $z\,{=}\,1$, gives rise to \textit{multiple zeta values} (MZVs):
\begin{equation}
	\label{eq:multizeta}
	(-1)^rG(\underbrace{0,\ldots,0,1}_{n_r},\underbrace{0,\ldots,0,1}_{n_{r-1}},\ldots,\underbrace{0,\ldots,0,1}_{n_1};1)=\zm(n_1,\ldots,n_r)=\sum_{0<k_1<\ldots<k_r}\frac{1}{k_1^{n_1}\cdots k_r^{n_r}}\,,
\end{equation}
with $r,n_1,\ldots,n_r\in\mathbb{N}$ and $n_r\geq2$. Shuffle relations carry over to MZVs from \eqn{eqn:shuffle1}, i.e.~for words $u,v\in X^\times$ one finds\footnote{For a word $w=x_0^{n_r-1}x_1x_0^{n_{r-1}-1}x_1\ldots x_0^{n_1-1}x_1\in X^\times$ we use the notation $\zm(n_1,\ldots,n_r)=\zm_w$.}
\begin{equation}
	\label{eqn:MZVshuffle}
	\zm_u\zm_v=\zm_{u\,\shuffle\,v}:=\sum_{w\in u\,\shuffle\,v}\zm_w\,, 
\end{equation}
where shuffle regularization is implemented by defining $\zeta_{x_1}=G(1;1)=0$\,.
\paragraph{Single-valued polylogarithms.} Multiple polylogarithms, as defined in \eqn{eqn:polylogdef}, are multi-valued functions in the complex plane. However, as shown in \rcite{BrownSVHPL}, one can combine polylogarithms of holomorphic and antiholomorphic arguments $z$ and $\zb$ in a way such that all branch cuts cancel. The resulting class of meromorphic functions are \textit{single-valued multiple polylogarithms} (svMPLs). The particular combinations yielding svMPLs can be identified from demanding trivial monodromy around the poles at zero and one \cite{Dixon:2012yy,BrownSVHPL}; the algorithm is spelled out in \appref{app:conSVHPL}. 

The class of single-valued polylogarithms relevant for this article can again be labeled by a two-letter alphabet, where each single-valued polylogarithm $\CG_w$ is again labeled by a word $w\in X^\times$. Defining furthermore the generating series
\begin{equation}
	\label{eqn:svmpl}
	\CG(z)=\sum_{w\in X^\times}\CG_w(z)\,w\,,
\end{equation}
one finds its holomorphic derivative to read
\begin{equation}
\label{eq:svhpl_gen_func}
\frac{\partial}{\partial z}\CG(z)=\left(\frac{x_0}{z}+\frac{x_1}{z-1}\right)\CG(z)\,,
\end{equation}
whereas the derivative with respect to the antiholomorphic variable $\zb$ will take a convenient form only when expressed in terms of a different alphabet \cite{BrownSVHPL}, as reviewed in appendix \ref{app:conSVHPL}. For the first examples of svMPLs one finds\footnote{In our definition of the svMPLs we have been starting from multiple polylogarithms $G$ in \eqn{eqn:polylogdef}. In \rcite{BrownSVHPL}, the construction of single-valued harmonic polylogarithms (svHPLs) is done using the harmonic polylogarithms $H$ as defined in \cite{Remiddi:2000}. Given relation \eqn{eqn:polylogrelations}, some of our svMPLs $\CG$ pick up a minus sign compared to the svHPLs $\CL$ defined by Brown in \rcite{BrownSVHPL}. The conversion is done by the map $x_1\mto -x_1$, which leads to $\CG_w=(-1)^{\#(x_1,w)}\CL_w$.}:
\begin{subequations}
\begin{align}
	\CG_0(z)&=G_0+\overline{G}_0, & \CG_1(z)&=G_1+\overline{G}_1,\\
	\CG_{00}(z)&=G_{00}+\overline{G}_{00}+G_0\overline{G}_0,& \CG_{01}(z)&=G_{01}+\overline{G}_{10}+G_0\overline{G}_1,\\
	\CG_{10}(z)&=G_{10}+\overline{G}_{01}+G_1\overline{G}_0,& \CG_{11}(z)&=G_{11}+\overline{G}_{11}+G_1\overline{G}_1.
\end{align}
\end{subequations}

\paragraph{Single-valued zeta values.} Corresponding to the MZVs $\zm(\vec{n})$, which were defined through multiple polylogarithms in \eqn{eq:multizeta}, so-called \textit{single-valued multiple zeta values} $\zm_{\text{sv}}(\vec{n})$ (svMZVs) are obtained as the corresponding special values of svMPLs. The algebra of svMZVs is well explored and understood \cite{Brown:2013gia}: using the labeling in terms of words, svMZVs abide the same shuffle relations \eqn{eqn:MZVshuffle} as their standard cousins. In a motivic framework, where the problem of proving linear independence of MZVs over $\ZQ$ can be addressed, svMZVs can be expressed as polynomials of common MZVs. 

\subsection{String amplitudes and the KLT relations}
\label{sec:klt}

\subsubsection{String amplitudes}
Scattering amplitudes in open and closed string theory are calculated as correlation functions of vertex operators inserted at the boundary (open string) or in the bulk (closed string) of a Riemann surface of appropriate genus. At tree level, the Riemann surfaces (or worldsheets) for open- and closed-string amplitudes are the disk and the Riemann sphere, respectively. 

In the notation of \rcite{KLT}, the formula for the bosonic $N$-point tachyon tree-level open-string amplitude reads
\begin{equation}
	\label{eqn:openstringamp}
	A_{\text{open}}^{(N)}(1,\ldots,N)=\frac{1}{\operatorname{vol}\SL(2,\mathbb{R})}\int_{-\infty}^\infty dx_1\cdots dx_N \prod_{x_i<x_j}(x_j-x_i)^{2\alpha'p_i\cdot p_j},
\end{equation}
where $\ap$ is the open-string inverse string tension and three integration variables can be fixed to cancel the volume factor of the Möbius group $\SL(2,\mathbb{R})$.
While the variables $x_i\in\ZR$, $i\in\{1,..,N\}$, are the (ordered) insertion points of the open-string vertex operators on the boundary of the disk, $p_i$, $i\in\{1\ldots N\}$, denote the momenta of the external strings. In general, when evaluating the string corrections of an open-string amplitude, all coefficients turn out to be standard MZVs defined in \eqn{eq:multizeta}. 

Analogously, an $N$-point tachyon tree-level bosonic closed-string amplitude reads (with the string coupling $\kappa$)
\begin{equation}
	\label{eqn:closedstringamp}
	A_{\text{closed}}^{(N)}=\frac{\pi}{\operatorname{vol}\SL(2,\mathbb{C})}\kappa^{N-2}\int d^2z_1\cdots d^2z_N \prod_{i<j}|z_i-z_j|^{\alpha'p_i\cdot p_j},
\end{equation}
where $\ap$ is the closed-string inverse string tension and again three integrations can be fixed due to Möbius invariance to cancel the volume of the symmetry group $\SL(2,\mathbb{C})$. On the Riemann sphere, vertex operator insertions for closed-string scattering amplitudes are complex variables $z_i\in\mathbb{C}$. For string-corrections to the closed scattering amplitudes, coefficients are svMZVs exclusively.

\begin{figure}[t]
\begin{center}
\scalebox{0.9}{
	\begin{tikzpicture}
				\shade[ball color = gray!40, opacity = 0.4] (-3.9,-0.2) arc (0:360:2);
				\draw (-3.9,-0.2) arc (0:-360:2);
				\draw (-3.9,-0.2) arc (0:-180:2 and 0.6);
				\draw[densely dashed] (-7.9,-0.2) arc (180:0:2 and 0.6);
				
				\draw[thick] (-4.7,0.5) -- (-4.5,0.7);
				\draw[thick] (-4.7,0.7) -- (-4.5,0.5);
				\draw[thick] (-6.6,0) -- (-6.4,0.2);
				\draw[thick] (-6.6,0.2) -- (-6.4,0);            	
				\draw[thick] (-5.2,-1) -- (-5.0,-1.2);
				\draw[thick] (-5.2,-1.2) -- (-5.0,-1);
				\draw[thick] (-6.8,-1.2) -- (-6.6,-1.4);
				\draw[thick] (-6.8,-1.4) -- (-6.6,-1.2);

        		\draw[very thick,->] (-3.65,-0.2) -- (-2.25,-0.2);
        	
        		\shade[ball color = gray!40, opacity = 0.4] (2,0) arc (0:180:2) arc (180:232:2 and 0.6) arc (174:-18:0.12) arc (241:307:2 and 0.6) arc (200:8:0.12) arc (315:360:2 and 0.6);
            	\shade[ball color = gray!40, opacity = 0.4] (2,-0.4) arc (0:-180:2) arc (180:128:2 and 0.6) arc (-174:16:0.12) arc (120:54.4:2 and 0.6) arc (-192:-13:0.12) arc (45:2:2 and 0.6); 
            	
            	\draw (2,0) arc (0:180:2);
            	\draw (-1.,-0.52) arc (-20:174:0.12);
            	\draw (1.403,-0.421) arc (10:202.05:0.12);
            	\draw[densely dashed] (+1.4,+0.42) arc (-20:174:0.12);
            	\draw[densely dashed] (-1.,+0.52) arc (6:200:0.12);
            	\draw (-2,0) arc (180:232.15:2 and 0.6); 
            	\draw (-1,-0.515) arc (240:306.09:2 and 0.6);
            	\draw (1.396,-0.42) arc (314.265:360:2 and 0.6);
            	\draw[dashed] (2,0) arc (0:45:2 and 0.6); 
				\draw[dashed] (1.17,0.48) arc (54:120:2 and 0.6);
				\draw[dashed] (-1.23,0.47) arc (128:180:2 and 0.6);        
        
            	\draw (2,-0.4) arc (0:-180:2);
            	\draw (-1.,-0.91) arc (-10:-191.6:0.12);
            	\draw (1.4,-0.83) arc (20:-172:0.12);
            	\draw[densely dashed] (+1.4,+0.038) arc (-10:-192:0.12);
            	\draw[densely dashed] (-1.,+0.12) arc (20:-172:0.12);
            	\draw (-1.89,-0.2) arc (161:182:2 and 0.6); 
            	\draw (-2,-0.4) arc (180:232:2 and 0.6);
            	\draw (-1,-0.915) arc (240:306:2 and 0.6);
            	\draw (1.395,-0.83) arc (314.2:380:2 and 0.6);
            	\draw[dashed] (2,-0.4) arc (0:45:2 and 0.6); 
            	\draw[dashed] (1.17,0.083) arc (54:120:2 and 0.6);
            	\draw[dashed] (-1.23,0.071) arc (127.6:161:2 and 0.6);

           		\draw[very thick,->] (2.25,-0.2) -- (3.65,-0.2); 

            	\fill[gray!40] (7.4,1.6) arc (0:40:1.7) arc (-45:-225:0.1491) arc (50:130:1.7) arc (45:-135:0.1491) arc (140:220:1.7) arc (135:-45:0.1491) arc (230:310:1.7) arc (-135:-315:0.1491) arc (320:360:1.7);
            	\fill[gray!40] (7.4,-2) arc (0:40:1.7) arc (-45:-225:0.1491) arc (50:130:1.7) arc (45:-135:0.1491) arc (140:220:1.7) arc (135:-45:0.1491) arc (230:310:1.7) arc (-135:-315:0.1491) arc (320:360:1.7);
            	
            	\draw (7.4,1.6) arc (0:40:1.7) arc (-45:-225:0.1491) arc (50:130:1.7) arc (45:-135:0.1491) arc (140:220:1.7) arc (135:-45:0.1491) arc (230:310:1.7) arc (-135:-315:0.1491) arc (320:360:1.7);
            	\draw (7.4,-2)arc (0:40:1.7) arc (-45:-225:0.1491) arc (50:130:1.7) arc (45:-135:0.1491) arc (140:220:1.7) arc (135:-45:0.1491) arc (230:310:1.7) arc (-135:-315:0.1491) arc (320:360:1.7);
            	
            	          
	\end{tikzpicture}
}
\caption{Graphical representation of the four-point KLT relation. The four-point closed-string amplitude is an integral over the Riemann sphere with four insertions of vertex operators. Cutting this sphere into two half-spheres with the insertions on the boundaries, one can rewrite the two contributions into two open-string four-point amplitudes.}
\label{fig:sphere}
\end{center}
\end{figure}
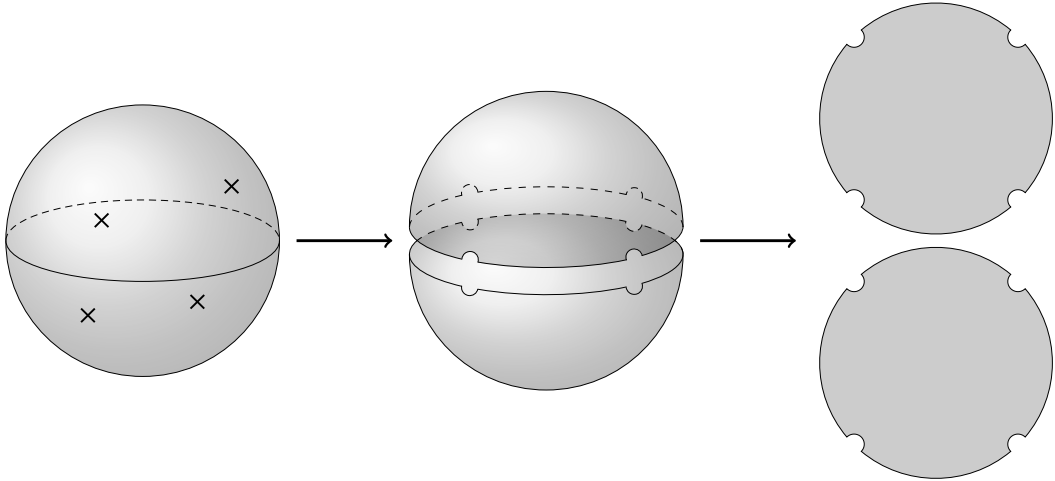

The Kawai--Lewellen--Tye relations will connect the two types of amplitudes in \eqns{eqn:openstringamp}{eqn:closedstringamp} as explained in the next subsection. For the four-point case this relation is depicted in \figref{fig:sphere}, indicating that the sphere with four punctures can be decomposed into two disks with each four vertices on the boundary.

\subsubsection{Relations between tree-level open- and closed-string scattering amplitudes}
\textit{Kawai--Lewellen--Tye (KLT) relations} \cite{KLT} are relations between tree-level scattering amplitudes of open and closed strings of equal number of insertions. They allow to express closed-string amplitudes at tree-level as sums of products of open-string tree-level amplitudes. Not surprisingly, they are compatible with the relations between standard MZVs and svMZVs as described in \rcites{Stieberger:2014hba,BrownDupont1}.

Again following the conventions used in \rcite{KLT}, we fix $\alpha'_{\text{open}}=\frac12$ and $\alpha'_\text{closed}=2$. The derivation of the KLT relations will be sketched for the case of four points, whereas the generalization to more external states can be found in the original reference.
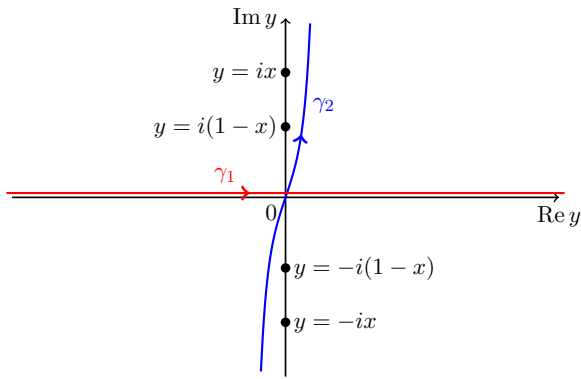
\begin{figure}
	\begin{center}
		\scalebox{0.8}{
\begin{tikzpicture}[decoration={markings,
mark=at position 4cm with {\arrow[line width=1.5pt]{>}}},
scale=0.9]
\draw[->, thick] (-5,0) -- (5,0) coordinate (xaxis);
\draw[->, thick] (0,-3.3) -- (0,3.3) coordinate (yaxis);

\fill (0,1.3)  circle[radius=2.5pt] node[left] {$y=i(1-x)$};
\fill (0,2.3)  circle[radius=2.5pt] node[left] {$y=ix$};

\fill (0,-1.3)  circle[radius=2.5pt] node[right] {$y=-i(1-x)$};
\fill (0,-2.3)  circle[radius=2.5pt] node[right] {$y=-ix$};

\draw[line width=1pt,postaction=decorate,domain=-0.45:0.45, smooth, variable=\x, blue] plot (\x, 3*\x+27*\x^3/3+27*18/15 *\x^5+118*\x^7);

\path[red,draw,line width=1pt,postaction=decorate](-5.1,0.08) -- (5.1,0.08);

\node[below] at (xaxis) {$\operatorname{Re} y$};
\node[left] at (yaxis) {$\operatorname{Im} y$};
\node[below left] {$0$};
\node[red] at (-1.1,0.4) {$\gamma_{1}$};
\node[blue] at (0.7,1.7) {$\gamma_{2}$};
\end{tikzpicture}
}
		\caption{Branch points and integration contour in the complex $y=\operatorname{Im}z$ plane. The branch points are represented by the dots. The red contour $\gamma_1$ represents the initial integration path for $y$ along the real axis. Noting the positions of the branch points it is possible to deform the contour for $y$ to the blue path $\gamma_2$. Figure reproduced from \rcite{KLT}.}
		\label{fig:branch_contour}
	\end{center}
\end{figure}
The four-point tachyon closed bosonic string amplitude (cf.~\eqn{eqn:closedstringamp} for $N=4$), with $z_1=0$, $z_3=1$, $z_4=\infty$ fixed due to the Möbius invariance, takes the form
\begin{equation}
\label{eq:amp4}
	A_{\text{closed}}^{(4)}=\pi \kappa^2\int d^2z |z|^{2p_{1}\cdot p_2}|1-z|^{2p_{2}\cdot p_3},
\end{equation}
where $d^2z=d\operatorname{Re}(z)\wedge d\operatorname{Im}(z)=\frac1{2i}d\zb\wedge dz$. Writing $z=x+iy$, for $x,y\in\mathbb{R}$, the integrand of the amplitude \eqn{eq:amp4} has four branch points in the complex $y$-plane at $y\in\{\pm ix,\pm i(1-x)\}$. These branch points are shown as the dots in \figref{fig:branch_contour} for $x\in[0,1]$. \Figref{fig:branch_contour} also shows the original integration contour $\gamma_1$ for $y$ along the real axis. Respecting the positions of the branch points in the complex $y$-plane allows to deform the integration contour to the path $\gamma_2$ along the imaginary axis shown in \figref{fig:branch_contour}. 
Introducing\footnote{The original article \cite{KLT} uses the variables $\xi$, $\eta$ instead of $\rho_\pm$.} $\rho_+{:=}\,x+iy$ and $\rho_-{:=}\,x-iy$, which are real after the deformation of the $y$-contour, and using the function
\begin{equation}
\label{eqn:funcf}
	f(p_{i}\cdot p_j;\rho_+,\rho_-):=\begin{cases} 0, & \text{for } \rho_+\rho_->0,\\p_{i}\cdot p_j, &\text{for } \rho_+\rho_-<0,\end{cases}
\end{equation}
to select the right branch of the integrand, the amplitude \eqn{eq:amp4} becomes
\begin{align}
A_{\text{closed}}^{(4)}=\frac{i}2\pi \kappa^2\int_{-\infty}^\infty d\rho_+\int_{-\infty}^\infty d\rho_- &|\rho_+|^{p_1\cdot p_2}|1-\rho_+|^{p_{2}\cdot p_3}|\rho_-|^{p_1\cdot p_2}|1-\rho_-|^{p_{2}\cdot p_3}\notag\\&\times e^{i\pi f(p_{1}\cdot p_2;\rho_+,\rho_-)+i\pi f(p_{2}\cdot p_3;(1-\rho_+),(1-\rho_-))}\,.
\end{align}%
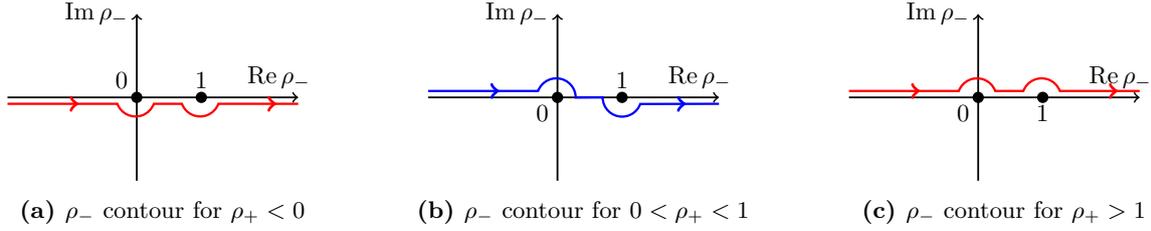
\begin{figure}
	\centering
	     \begin{subfigure}[b]{0.3\textwidth}
         \centering
         \scalebox{0.85}{
         \begin{tikzpicture}[decoration={markings,
		mark=at position 1.1cm with {\arrow[line width=1.5pt]{>}},
		mark=at position 4.5cm with {\arrow[line width=1.5pt]{>}}
		}
		]
		\draw[->, thick] (-2,0) -- (2.5,0) coordinate (xaxis);
		\draw[->, thick] (0,-1.3) -- (0,1.3) coordinate (yaxis);

		\fill (0,0)  circle[radius=2.5pt] node[above left] {$0$};
		\fill (1,0)  circle[radius=2.5pt] node[above] {$1$};
		\path[red,draw,line width=1pt,postaction=decorate] (-2,-0.1) -- (-0.3,-0.1) arc (200:340:0.3) -- (0.7,-0.1) arc (200:340:0.3) -- (2.5,-0.1);

		\node[above] at (2.2,0) {$\operatorname{Re} \rho_-$};
		\node[left] at (yaxis) {$\operatorname{Im} \rho_-$};
		\end{tikzpicture}
		}
         \caption{$\rho_-$ contour for $\rho_+<0$}
         \label{fig:contours1}
     \end{subfigure}
     \hfill
     \begin{subfigure}[b]{0.3\textwidth}
         \centering
         \scalebox{0.85}{
         \begin{tikzpicture}[decoration={markings,
		mark=at position 1.1cm with {\arrow[line width=1.5pt]{>}},
		mark=at position 4.5cm with {\arrow[line width=1.5pt]{>}}
		}
		]
		\draw[->, thick] (-2,0) -- (2.5,0) coordinate (xaxis);
		\draw[->, thick] (0,-1.3) -- (0,1.3) coordinate (yaxis);

		\fill (0,0)  circle[radius=2.5pt] node[below left] {$0$};
		\fill (1,0)  circle[radius=2.5pt] node[above] {$1$};
		\path[blue,draw,line width=1pt,postaction=decorate] (-2,0.1) -- (-0.3,0.1) arc (160:0:0.3) -- (0.7,0) arc (180:340:0.3) -- (2.5,-0.1);

		\node[above] at (2.2,0) {$\operatorname{Re} \rho_-$};
		\node[left] at (yaxis) {$\operatorname{Im} \rho_-$};
		\end{tikzpicture}
		}
         \caption{$\rho_-$ contour for $0<\rho_+<1$}
         \label{fig:contours2}
     \end{subfigure}
     \hfill
     \begin{subfigure}[b]{0.3\textwidth}
         \centering
         \scalebox{0.85}{
          \begin{tikzpicture}[decoration={markings,
		mark=at position 1.1cm with {\arrow[line width=1.5pt]{>}},
		mark=at position 4.5cm with {\arrow[line width=1.5pt]{>}}
		}
		]
		\draw[->, thick] (-2,0) -- (2.5,0) coordinate (xaxis);
		\draw[->, thick] (0,-1.3) -- (0,1.3) coordinate (yaxis);

		\fill (0,0)  circle[radius=2.5pt] node[below left] {$0$};
		\fill (1,0)  circle[radius=2.5pt] node[below] {$1$};
		\path[red,draw,line width=1pt,postaction=decorate] (-2,0.1) -- (-0.3,0.1) arc (160:20:0.3) -- (0.7,0.1) arc (160:20:0.3) -- (2.5,0.1);

		\node[above] at (2.2,0) {$\operatorname{Re} \rho_-$};
		\node[left] at (yaxis) {$\operatorname{Im} \rho_-$};
		\end{tikzpicture}
		}
         \caption{$\rho_-$ contour for $\rho_+>1$}
         \label{fig:contours3}
     \end{subfigure}
	\caption{Integration contours for $\rho_-$ for different values of $\rho_+$. In the cases (a) and (c) the contour can be closed at infinity below and above the real axis, respectively, such that the integrals vanish using Cauchy's theorem. Thus, only the contour (b) delivers a contribution. 
	Figures are reproduced from \rcite{KLT}.}
	\label{fig:contours}
\end{figure}%
In order to separate the two integrations in $\rho_+$ and $\rho_-$, three different integration paths for $\rho_-$ are considered, depending on the value of $\rho_+$. These are the three cases $\rho_+{<}\,0$, $0\,{<}\,\rho_+{<}\,1$ and $\rho_+{>}\,1$, which are all depicted in \figref{fig:contours}. For the cases $\rho_+{<}\,0$ (\figref{fig:contours1}) and $\rho_+{>}\,1$ (\figref{fig:contours3}), the integral contours can be closed at infinity in the lower and upper complex plane, respectively, implying that the integrals vanish due to Cauchy's residue theorem. Only the part of the integral with $0\,{<}\,\rho_+{<}\,1$ remains (\figref{fig:contours2}). This contour is deformed as shown in \figref{fig:kltcontour}, which -- after choosing the right phase factors for the integrand -- finally disentangles the two integrations of $\rho_+$ and $\rho_-$ completely. 

The resulting four-point closed-string tachyon amplitude is
\begin{equation}
\label{eq:closed4}
	A_{\text{closed}}^{(4)}=-\pi \kappa^2\sin(\pi p_{2}\cdot p_3)\int_0^1d\rho_+ |\rho_+|^{p_{1}\cdot p_2}|1-\rho_+|^{p_{2}\cdot p_3}\int_1^\infty d\rho_- |\rho_-|^{p_{1}\cdot p_2}|1-\rho_-|^{p_{2}\cdot p_3}.
\end{equation}%
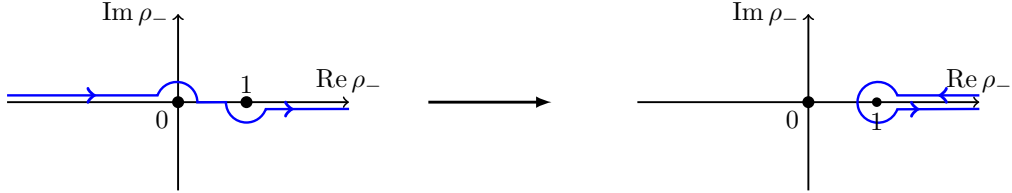
\begin{figure}
    \centering
    \scalebox{0.9}{     \begin{subfigure}[b]{0.4\textwidth}
         \centering
         \begin{tikzpicture}[decoration={markings,
		mark=at position 1.3cm with {\arrow[line width=1.5pt]{>}},
		mark=at position 4.7cm with {\arrow[line width=1.5pt]{>}}
		}
		]
		\draw[->, thick] (-2.5,0) -- (2.5,0) coordinate (xaxis);
		\draw[->, thick] (0,-1.3) -- (0,1.3) coordinate (yaxis);

		\fill (0,0)  circle[radius=2.5pt] node[below left] {$0$};
		\fill (1,0)  circle[radius=2.5pt] node[above] {$1$};
		\path[blue,draw,line width=1pt,postaction=decorate] (-2.5,0.1) -- (-0.3,0.1) arc (160:0:0.3) -- (0.7,0) arc (180:340:0.3) -- (2.5,-0.1);

		\node[above] at (xaxis) {$\operatorname{Re} \rho_-$};
		\node[left] at (yaxis) {$\operatorname{Im} \rho_-$};
	\end{tikzpicture}
     \end{subfigure}
     \hfill
     \begin{subfigure}[b]{0.16\textwidth}
         \centering
		\begin{tikzpicture}
		\draw[-latex, very thick] (3,1.3) -- (4.8,1.3);
		\draw[transparent,-latex, thick] (3,0) -- (5.5,0);
		\end{tikzpicture}
     \end{subfigure}
     \hfill
     \begin{subfigure}[b]{0.4\textwidth}
         \centering
         \begin{tikzpicture}[decoration={markings,
		mark=at position 0.6cm with {\arrow[line width=1.5pt]{>}},
		mark=at position 3.2cm with {\arrow[line width=1.5pt]{>}}
		}
		]
		\draw[->, thick] (-2.5,0) -- (2.5,0) coordinate (xaxis);
		\draw[->, thick] (0,-1.3) -- (0,1.3) coordinate (yaxis);

		\fill (0,0)  circle[radius=2.5pt] node[below left] {$0$};
		\fill (1,0)  circle[radius=2pt];
		\path[blue,draw,line width=1pt,postaction=decorate] (2.5,0.1) -- (1.3,0.1) arc (20:340:0.3) --  (2.5,-0.1);

		\node[above] at (xaxis) {$\operatorname{Re} \rho_-$};
		\node[left] at (yaxis) {$\operatorname{Im} \rho_-$};
		\node[above] at (1,-0.55) {$1$};
	\end{tikzpicture}
     \end{subfigure}}
    \caption{Deformation of the $\rho_-$ integration contour for the case $0\,{<}\,\rho_+{<}\,1$. This step disentangles the two integration variables $\rho_-$ and $\rho_+$ and the integration domains become $\rho_-{\in}[1,\infty)$ and $\rho_+{\in}[0,1]$. Figure taken from \rcite{KLT}.}
        \label{fig:kltcontour}
\end{figure}%
Considering the two integrals on the right-hand side of \eqn{eq:closed4}, one finds the first integral resembling an open four-point tachyon amplitude with ordering $(1,2,3,4)$ and the second integral the corresponding amplitude with the ordering $(1,3,2,4)$. Thus, one finds the \textit{KLT relation for the four-point tachyon amplitude}
\begin{equation}
	A_{\text{closed}}^{(4)}=-\pi \kappa^2\sin(\pi p_{2}\cdot p_3)A_{\text{open}}^{(4)}(1,2,3,4)A_{\text{open}}^{(4)}(1,3,2,4)\,,
\end{equation}
which combines two open-string amplitudes with a kinematic sine factor into the closed-string tachyon amplitude. 

As spelled out in \cite{KLT}, the result can be generalized to higher-point amplitudes and reads for the $N$-point tachyon amplitude
\begin{equation}
	A_{\text{closed}}^{(N)}=\left(\frac{i}2\right)^{N-3}\pi \kappa^{N-2}\sum_{P,P'}\overline{A}_{\text{open}}^{(N)}(P)A_{\text{open}}^{(N)}(P')e^{i\pi F(P,P')}.
\end{equation}
The phase factor in this formula is similar to that of the four-point case but takes into account the orderings $P$ and $P'$ of the $N-3$ integration variables $\rho_+^{(2)},\ldots,\rho_+^{(N-2)}$ and $\rho_-^{(2)},\ldots,\rho_-^{(N-2)}$, respectively, through the function $F$, a generalization of the function $f$ in \eqn{eqn:funcf}. In other words, the single-valued closed-string amplitude (which is an integral over the punctured Riemann sphere) can be written as a sum of products of (multi-valued) open-string amplitudes (defined on the disk).


\section{\texorpdfstring{Multi-Regge kinematics in \Nfour sYM theory}{Multi-Regge kinematics in N=4 sYM theory}}
\label{sec:BFKLMRK}
In this section, we briefly review gluon scattering in the multi-Regge limit in \Nfour sYM theory. This is done in preparation for \secref{sec:doublecopy}, where we are going to show that the six-point remainder function in planar \Nfour sYM theory can be decomposed into two contributions in a way very similar to the decomposition of the four-point closed-string tachyon amplitude reviewed in \subsecref{sec:klt}.

The interest in this scenario originates from the fact that gluon scattering in multi-Regge kinematics (MRK) is governed by the Balitsky--Fadin--Kuraev--Lipatov (BFKL) differential equation \cite{Fadin:1975cb, Kuraev:1976ge, Kuraev:1977fs, Balitsky:1978ic}, for which recursive solutions are readily available. The multi-Regge limit leads to a decreased number of degrees of freedom in the kinematical space and thereby enhances the overall symmetry of the problem. Effectively, the increased symmetry can be shown to translate into the appearance of single-valued polylogarithms and associated single-valued zeta values exclusively \cite{Dixon:2012yy} in the remainder functions. With this feature, MRK amplitudes resemble closed-string scattering amplitudes: this triggered the investigation reported on in this article. As will be discussed below, the simplest object deviating from the BDS ansatz in MRK is the six-point amplitude. Accordingly, this is the point where one will most likely be able to identify an analogue of the KLT decomposition of the four-point closed-string amplitude. 

The following review is mostly based on \rcite{Del_Duca_2016}, whereas a general review on scattering amplitudes in MRK can be found in ref.~\cite{DelDucaDixon}. 

\subsection{Multi-Regge kinematics} 
Let us consider an $N$-particle scattering process, where two incoming particles are scattered into $N-2$ outgoing particles. External momenta are denoted by $p_1$, $p_2$ for the incoming particles and $p_3,\ldots,p_N$ for the outgoing particles, with components $(p_i^0,p_i^1,p_i^2,p_i^3)$, $i=1,\ldots,N$. In order to describe the multi-Regge situation, it is useful to use light-cone momenta 
\begin{equation}
	p^\pm_k:=p^0_k\pm p^3_k
\end{equation}
and complex transverse momenta
\begin{equation}
\label{eq:comp_trans_mom}
p^c_k:=p_{k\perp}=p^1_k+ip^2_k\,.
\end{equation}
Choosing a center-of-mass Lorentz frame such that $p_2^3=p_2^0$, the lightcone momenta $p_1^+$ and $p_2^-$ vanish, as well as the transverse momenta of the incoming particles:
\begin{equation}
	p_1^+=p_2^-=p^c_1=p^c_2=0\,.
\end{equation}
Including the above choice of the Lorentz frame, the multi-Regge limit is defined as the regime where the rapidities (or equivalently in the light-cone plus coordinates) of the outgoing particles are strongly ordered, while transverse momenta are assumed to be comparable:
\begin{equation}
p_3^+\gg p_4^+\gg\ldots\gg p_{N-1}^+\gg p_N^+\,,\quad |p^c_3|\simeq |p^c_4|\simeq\ldots\simeq |p^c_N|\,.
\end{equation}
This limit can be understood as a head-to-head collision of particles 1 and 2 in the center-of-mass-frame with just a little transverse offset of the momenta.

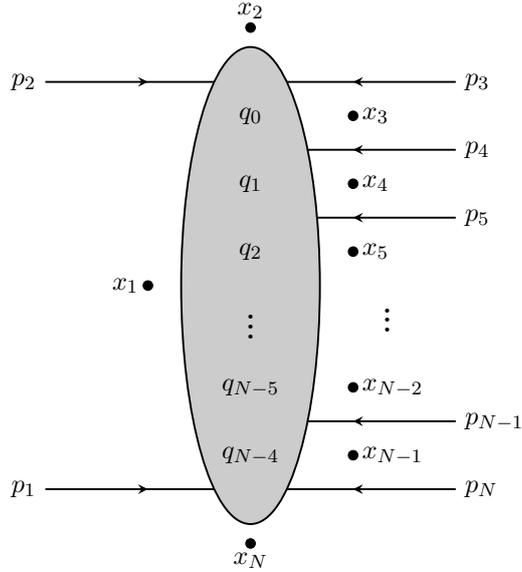
\begin{figure}[t]
	\centering
	\scalebox{0.9}{
	\begin{tikzpicture}

\node[left] at (-3,3) {$p_2$};
\draw[thick,-stealth] (-3,3) -- (-1.5,3);
\draw[thick] (-1.6,3) -- (0,3) ;
\draw[thick,-stealth] (3,3) -- (1.5,3);
\draw[thick] (1.6,3) -- (0,3);
\node[left] at (-3,-3) {$p_1$};
\draw[thick,-stealth] (-3,-3) -- (-1.5,-3);
\draw[thick] (-1.6,-3) -- (0,-3);
\draw[thick,-stealth] (3,-3) -- (1.5,-3);
\draw[thick] (1.6,-3) -- (0,-3);
\node[right] at (3,3) {$p_3$};
\node[right] at (3,2) {$p_4$};
\node[right] at (3,1) {$p_5$};
\node[right] at (3,-2) {$p_{N-1}$};
\node[right] at (3,-3) {$p_N$};
\node at (2,-0.4) {\huge\vdots};
\draw[thick] (0,2) -- (1.6,2);
\draw[thick,stealth-] (1.5,2) -- (3,2);
\draw[thick] (0,1) -- (1.6,1);
\draw[thick,stealth-] (1.5,1) -- (3,1);
\draw[thick] (0,-2) -- (1.6,-2);
\draw[thick,stealth-] (1.5,-2) -- (3,-2);
\filldraw[black] (0,3.8) circle (2pt) node[above]{$x_2$};
\filldraw[black] (0,-3.8) circle (2pt) node[below]{$x_N$};
\filldraw[black] (1.5,2.5) circle (2pt) node[right]{$x_3$};
\filldraw[black] (1.5,1.5) circle (2pt) node[right]{$x_4$};
\filldraw[black] (1.5,0.5) circle (2pt) node[right]{$x_5$};
\filldraw[black] (1.5,-1.5) circle (2pt) node[right]{$x_{N-2}$};
\filldraw[black] (1.5,-2.5) circle (2pt) node[right]{$x_{N-1}$};
\filldraw[black] (-1.5,0) circle (2pt) node[left]{$x_1$};

\draw[ultra thick] (0,0) ellipse (1 and 3.5);
\fill[gray!40] (0,0) ellipse (1 and 3.5);

\node at (0,2.5) {$q_0$};
\node at (0,1.5) {$q_1$};
\node at (0,0.5) {$q_2$};
\node at (0,-0.5) {\huge\vdots};
\node at (0,-1.5) {$q_{N-5}$};
\node at (0,-2.5) {$q_{N-4}$};

\end{tikzpicture}
	}
	\caption{Kinematical variables in the multi-Regge regime. All particles are considered incoming.}
	\label{fig:MRK_variables}
\end{figure}

For convenience let us define generalized Mandelstam variables
\begin{equation}
\label{eq:mandelvar}
s_{i(i+1)\ldots j}=(p_i+p_{i+1}+\ldots+p_j)^2=x_{(i-1)j}^2\,, \quad t_{i+1}=q_i^2\,, \quad q_i=-p_2-\ldots-p_{i+3}=x_{(i+3)1}\,.
\end{equation}
The variables $q_i$ denote the momenta transferred by the internal gluons of the process, while the $x_{ij}=x_i-x_j$ are differences of the dual variables $x_i$ where $i$ and $j$ are understood modulo $N$. Invariance of these variables under \textit{dual conformal symmetry} \cite{Drummond:2011} is instrumental in showing the single-valuedness of the remainder function \cite{Dixon:2012yy}. The dual variables $x_i$ are related to the external momenta by $p_i=x_i-x_{i-1}$. All variables involved in the description of the scattering process in MRK are displayed in \figref{fig:MRK_variables}.
In terms of generalized Mandelstam variables defined in \eqn{eq:mandelvar}, the multi-Regge limit can be expressed as 
\begin{align}
s_{12}\gg s_{3\ldots (N-1)},\,s_{4\ldots N}&\gg s_{3\ldots (N-2)},\,s_{4\ldots (N-1)},\, s_{5\ldots N}\gg\ldots\nonumber\\
&\ldots\gg s_{34},\,\ldots,\, s_{(N-1) N}\gg -t_1,\,\ldots ,\,-t_{N-3}\,.
\end{align}

\paragraph{Multi-Regge kinematics in $\CN{=}\,4$ sYM theory.} Specializing to scattering in planar $\CN{=}\,4$ sYM theory, one can make use of conformal cross-ratios $U_{ij}$, defined in \rcite{Drummond:2010}:
\begin{equation}
	\label{eqn:confcross}
U_{ij}=\frac{x_{(i+1)j}^2x_{i(j+1)}^2}{x_{ij}^2x_{(i+1)(j+1)}^2}\,,\quad 3\leq|i-j|<N-2\,.
\end{equation}
Not all of those cross-ratios are independent. Using momentum conservation, one can show that there are $3N-15$ independent cross-ratios for $N$-gluon scattering. 

\paragraph{Example: six-point scattering.}
For scattering of six gluons, there are three independent cross-ratios. In the multi-Regge limit, where $\delta=\frac{p_5^+}{p_4^+}\rightarrow0$, they can be expanded as \cite{Del_Duca_2016}
\begin{subequations}
\begin{align}
\label{eq:MRK_limit}
u_1&:=U_{36}=\frac{x_{13}^2 x_{46}^2}{x_{14}^2 x_{36}^2}=1-\delta\frac{|p^c_4+p^c_5|^2}{|p^c_5|^2}+\mathcal{O}(\delta^2)= 1+\mathcal{O}(\delta),\\
u_2&:=U_{14}=\frac{x_{24}^2x_{15}^2}{x_{25}^2x_{14}^2}=\delta\frac{|q^c_0|^2}{|q^c_1|^2}+\mathcal{O}(\delta^2)= 0+\mathcal{O}(\delta),\\
u_3&:=U_{25}=\frac{x_{35}^2x_{26}^2}{x_{36}^2x_{25}^2}=\delta\frac{|q^c_2|^2|p^c_4|^2}{|q^c_1|^2|p^c_5|^2}+\mathcal{O}(\delta^2)= 0+\mathcal{O}(\delta),
\end{align}
\end{subequations}
with $p^c_i$, $q^c_i$ defined in \eqn{eq:comp_trans_mom}. Finally, the \textit{reduced cross-ratios} 
\begin{equation}
	\label{eq:crossratios}
	\tilde{u}_2=\frac{u_2}{1-u_1}=\frac{1}{|1+w|^2}\,,\quad \tilde{u}_3=\frac{u_3}{1-u_1}=\frac{|w|^2}{|1+w|^2}
\end{equation}
remain finite when taking the multi-Regge limit \eqn{eq:MRK_limit}. The anharmonic ratios $w$ and $w^*$, implicitly defined in \eqn{eq:crossratios} above, together with the cross-ratio $u_1$ capture the kinematical freedom of the MRK limit. In other words, the six-point results in MRK can be expressed solely in terms of the three variables $u_1,w,w^*$. This set of variables is going to be used throughout the remainder of this article.

In the multi-Regge limit, scattering amplitudes turn out to have deep mathematical structures, which was already used in the 1970s in the BFKL equation \cite{Fadin:1975cb, Kuraev:1976ge, Kuraev:1977fs, Balitsky:1978ic} (originally linked to pomeron scattering in QCD). The BFKL approach uses resummation of the radiative corrections in orders of the logarithm of the ratio of the squared center-of-mass energy and the fixed transferred momentum of the process. This approach needs the MRK gluon amplitudes as key ingredients and thus relies on the MRK. In the gluon scattering amplitudes in \Nfour sYM theory which we will investigate in this article, the BFKL equation describes the internal gluon processes and yields the BFKL eigenvalue, which appears in the Mellin-integral of the remainder function for amplitudes with more than five gluons (see later parts of the following subsection). For four- and five-gluon scattering, the amplitude is completely described by the BDS ansatz \cite{Bern:2005iz}, which will be described next.

\subsection{BDS ansatz and six-point remainder function}
\label{sec:6pt}
Scattering of $N$ gluons in $\mathcal{N}{=}\,4$ sYM theory at any loop order $L$ was conjectured to be expressible in terms of the Bern--Dixon--Smirnov (BDS) ansatz \cite{Bern:2005iz}. It reads 
\begin{equation}
	\label{eq:BDSansatz}
	\CM_N^{\text{BDS}}=1+\sum_{L=1}^\infty a^LM_N^{(L)}(\varepsilon)=\exp\left[\,\sum_{\ell=1}^\infty a^\ell\left(f^{(\ell)}(\varepsilon)M_N^{(1)}(\ell\varepsilon)+C^{(\ell)}+E_N^{(\ell)}(\varepsilon)\right)\right], 
\end{equation}
where $M_N^{(L)}$ denotes the $L$-loop $N$-point amplitude to all orders in $\varepsilon$ (for the dimension $d=4-2\varepsilon$)
and each new loop order comes with an additional power of the coupling constant
\begin{equation}
	\label{eqn:BDSloopcounting}
	a=\frac{N_c\,\alpha_s}{2\pi}(4\pi e^{-\gamma_E})^\varepsilon\,.
\end{equation}
In the above equation, $N_c$ is the number of colors, $\alpha_s$ the strong-coupling constant and the Euler--Mascheroni constant is $\gamma_E=-\Gamma'(1)$. In the second part of \eqn{eq:BDSansatz}, the generating function $\CM_N^{\text{BDS}}$ is expressed as the weighted exponential of the one-loop contribution $M_N^{(1)}$ along with several correction terms. In particular, the functions $f^{(\ell)}(\varepsilon)$ are related to the cusp anomalous dimension and can be written as a second-order polynomial in $\varepsilon$:
\begin{equation}
	f^{(\ell)}(\varepsilon)=f_0^{(\ell)}+\varepsilon f_1^{(\ell)}+\varepsilon^2 f_2^{(\ell)}.
\end{equation}
The constants $f_k^{(\ell)}$ and $C^{(\ell)}$ are sums of zeta values multiplied by rational numbers, such that the principle of maximal transcendentality \cite{Kotikov:2003} is respected\footnote{Functions $f_k^{(\ell)}$ have the degree of transcendentality $2\ell-2+k$ and the constants $C^{(\ell)}$ have $2\ell$, respectively \cite{Bern:2005iz}.}. These constants are independent of the number of legs $N$ of the scattering amplitude in question. Finally, the functions $E_N^{(\ell)}(\varepsilon)$ are contributions vanishing in the limit $\varepsilon\rightarrow0$.

While the BDS ansatz for $N{=}\,4$ and $N{=}\,5$ gluons was shown to reproduce all known results~\cite{Drummond:2010}, starting from $N{=}\,6$ gluons, the BDS ansatz needs to be corrected by a factor $e^{R_N}$ (see e.g.~refs.~\cite{two-loop,BFKLPomeron,reggecut,BFKLapproach}):
\begin{equation}
	\label{eq:BDSdividedout}
	\CA_N=\CA_N^{\text{BDS}}e^{R_N}.
\end{equation}
In this equation, $\CA$ denotes the full amplitude $\CM$ divided by the tree amplitude. The function $R_N$ is called the \textit{remainder function} \cite{two-loop} and it is a function of the conformal cross-ratios defined in \eqn{eqn:confcross} and therefore exhibits dual conformal symmetry. For $N{=}\,4$ and $N{=}\,5$ there are no independent cross-ratios, and thus $R_4\,{=}\,R_5\,{=}\,0$ \cite{Drummond:2010}. A closed integral formula for the first non-trivial remainder function $R_6$ in MRK has been derived from the BFKL formalism\footnote{Another approach to obtain a closed form for the remainder function is via hexagonal Wilson loops~\cite{Alday:2007, Brandhuber:2008, Brandhuber:2010, Basso:2015}.} \cite{BFKLPomeron,reggecut,BFKLapproach}. In the next paragraphs we are going to review the main steps and arguments for the derivation of the remainder function.   

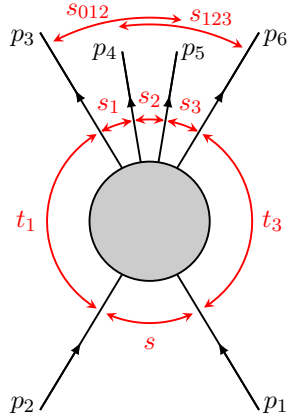
\begin{figure}
	\begin{center}
		\scalebox{0.9}{
		\begin{tikzpicture}

\draw[thick, fill=gray!40] (0,0) circle[radius=25pt];
\draw[thick,-latex] (-1.6,-2.78) -- (-1,-1.78);
\draw[thick] (-1.6,-2.78) -- (-0.4,-0.78);
\draw[thick,-latex] (1.6,-2.78) -- (1,-1.78);
\draw[thick] (1.6,-2.78) -- (0.4,-0.78);

\draw[thick,latex reversed-]  (-1,1.78) -- (-1.6,2.78);
\draw[thick] (-1.6,2.78) -- (-0.4,0.78);
\draw[thick,latex reversed-]  (1,1.78) -- (1.6,2.78);
\draw[thick] (1.6,2.78) -- (0.4,0.78);
\draw[thick,latex reversed-]  (-0.53/2,3.37/2) -- (-0.4,2.5);
\draw[thick] (-0.4,2.5) -- (-0.13,0.87);
\draw[thick,latex reversed-]  (0.53/2,3.37/2) -- (0.4,2.5);
\draw[thick] (0.4,2.5) -- (0.13,0.87);

\node at (1.85,-2.7) {$p_1$};
\node at (-1.85,-2.7) {$p_2$};
\node at (1.85,2.7) {$p_6$};
\node at (-1.85,2.7) {$p_3$};
\node at (-0.65,2.5) {$p_4$};
\node at (0.65,2.5) {$p_5$};

\draw [stealth-stealth,red,thick,domain=245:295] plot ({1.5 * cos(\x)}, {1.5 * sin(\x)});
\draw [stealth-stealth,red,thick,domain=302:419] plot ({1.5 * cos(\x)}, {1.5 * sin(\x)});
\draw [stealth-stealth,red,thick,domain=121:239] plot ({1.5 * cos(\x)}, {1.5 * sin(\x)});
\draw [stealth-stealth,red,thick,domain=100:118] plot ({1.5 * cos(\x)}, {1.5 * sin(\x)});
\draw [stealth-stealth,red,thick,domain=62:80] plot ({1.5 * cos(\x)}, {1.5 * sin(\x)});
\draw [stealth-stealth,red,thick,domain=82:98] plot ({1.5 * cos(\x)}, {1.5 * sin(\x)});
\draw [stealth-stealth,red,thick,domain=81:118] plot ({3 * cos(\x)}, {3 * sin(\x)});
\draw [stealth-stealth,red,thick,domain=62:99] plot ({2.9 * cos(\x)}, {2.9 * sin(\x)});

\node[red] at (0,-1.8) {$s$};
\node[red] at (1.8,0) {$t_3$};
\node[red] at (-1.8,0) {$t_1$};
\node[red] at (-0.6,1.7) {$s_1$};
\node[red] at (0,1.8) {$s_2$};
\node[red] at (0.6,1.7) {$s_3$};
\node[red] at (-0.9,3.1) {$s_{012}$};
\node[red] at (0.9,3) {$s_{123}$};

\end{tikzpicture}
		}
		\caption{Pictorial definitions of the relevant Mandelstam variables for the $2\rightarrow4$ gluon scattering. The two incoming lines at the bottom represent the two gluons before the scattering while the four outgoing lines at the top represent the gluons after the interaction. Figure adapted from ref.~\cite{reggecut}.
		}
		\label{fig:variables}
	\end{center}
\end{figure}

\paragraph{The remainder function $R_6$.} To calculate the six-gluon amplitude in MRK, we consider a scattering process as depicted in \figref{fig:variables}. The multi-Regge regime in the variables of \figref{fig:variables} reads
\begin{equation}
	s\gg s_1,s_2,s_3\gg t_1,t_2,t_3\,.
\end{equation}
For the $2\,{\rightarrow}\,4$ gluon scattering process, different configurations of Mandelstam variables exist, depending on the choice of sign for the Mandelstam variables: they are referred to as \textit{physical regions} and are related by analytical continuation in the dual conformal cross-ratios. An important subtlety when going from one region to another is that the corresponding analytic continuation needs to be performed \textit{before} taking the MRK limit. The scattering amplitude now might take different forms in the different physical regions, which may or may not obey the BDS ansatz.

\begin{figure}
	\begin{center}
		\scalebox{0.9}{
		\begin{tikzpicture}

\node at (-3.5,3) {(a)};
\node at (5.5,3) {(b)};

\node[left] at (-2,3) {$p_2$};
\draw[thick,-stealth] (-2,3) -- (-1,3);
\draw[thick] (-1.1,3) -- (0,3) ;
\draw[thick,-stealth reversed] (2,3) -- (1,3);
\draw[thick] (1.1,3) -- (0,3);
\node[left] at (-2,-1.5) {$p_1$};
\draw[thick,-stealth] (-2,-1.5) -- (-1,-1.5);
\draw[thick] (-1.1,-1.5) -- (0,-1.5);
\draw[thick,-stealth reversed] (2,-1.5) -- (1,-1.5);
\draw[thick] (1.1,-1.5) -- (0,-1.5);
\node[right] at (2,3) {$p_3$};
\node[right] at (2,1.5) {$p_4$};
\node[right] at (2,0) {$p_5$};
\node[right] at (2,-1.5) {$p_6$};
\draw[thick] (0,1.5) -- (1.1,1.5);
\draw[thick,stealth reversed-] (1,1.5) -- (2,1.5);
\draw[thick] (0,0) -- (1.1,0);
\draw[thick,stealth reversed-] (1,0) -- (2,0);

\draw[ultra thick] (0,0.75) ellipse (0.7 and 2.5);
\fill[gray!40] (0,0.75) ellipse (0.7 and 2.5);

\draw[-latex, thick] (3,0.5) -- (6,0.5);
\node at (4.5,1) {$u_1 e^{i0}\longrightarrow u_1e^{-2\pi i}$};

\node[left] at (7,3) {$p_2$};
\draw[thick,-stealth] (7,3) -- (8,3);
\draw[thick] (7.9,3) -- (9,3) ;
\draw[thick,-stealth reversed] (11,3) -- (10,3);
\draw[thick] (10.1,3) -- (9,3);
\node[left] at (7,-1.5) {$p_1$};
\draw[thick,-stealth] (7,-1.5) -- (8,-1.5);
\draw[thick] (7.9,-1.5) -- (9,-1.5);
\draw[thick,-stealth reversed] (11,-1.5) -- (10,-1.5);
\draw[thick] (10.1,-1.5) -- (9,-1.5);
\node[right] at (11,3) {$p_3$};
\node[left] at (7.4,2.1) {$p_4$};
\node[left] at (7.4,0.6) {$p_5$};
\node[right] at (11,-1.5) {$p_6$};

\draw[thick] (9,0) -- (10.2,0);
\draw[thick] (9,1.5) -- (10.2,1.5);
\draw[ultra thick] (9,0.75) ellipse (0.7 and 2.5);
\fill[gray!40] (9,0.75) ellipse (0.7 and 2.5);

\draw[thick] (10.2,0) arc (-90:90:0.3);
\draw[thick] (10.2,0.6) -- (7.4,0.6);
\draw[thick,stealth reversed-] (8.7,0.6) -- (7.4,0.6);

\draw[thick] (10.2,1.5) arc (-90:90:0.3);
\draw[thick] (10.2,2.1) -- (7.4,2.1);
\draw[thick,stealth reversed-] (8.7,2.1) -- (7.4,2.1);

\end{tikzpicture}
		}
		\caption{Analytic continuation from the Euclidean region (a) with only positive energies, to the Mandelstam region (b) where the deviation from the BDS ansatz becomes non-trivial. Figure adapted from \rcite{BFKLapproach}.}
		\label{fig:analyticcont}
	\end{center}
\end{figure}
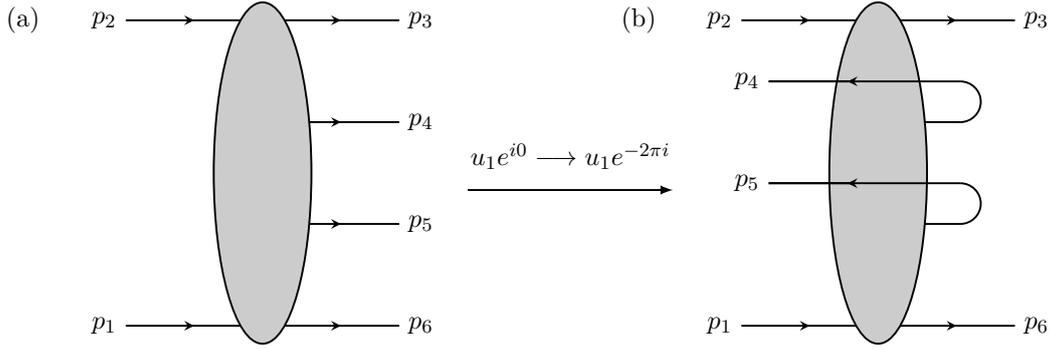

The first region under consideration is the so-called \textit{Euclidean region}, which is pictured on the left-hand side of \figref{fig:analyticcont}. In terms of Mandelstam variables it is defined by
\begin{equation}
	s,s_1,s_2,s_3,s_{012},s_{123}>0\,.
\end{equation}
In this region the amplitude takes the exact form of the BDS ansatz (see e.g.~ref.~\cite{BFKLPomeron}). Different regions can be reached from the Euclidean region by supplementing a phase to one or more of the dual conformal cross-ratios $u_i$. This analytic continuation is conveniently pictured by flipping sides of one or more of the central outgoing legs as shown in \figref{fig:analyticcont}. 

When only one of the central legs is flipped compared to the Euclidean region, the resulting amplitude will still obey the BDS ansatz \cite{Bern:2005iz}. A non-trivial situation breaking the BDS ansatz occurs when both of the centrally produced outgoing particles are flipped. This scenario is depicted in \figref{fig:analyticcont}(b) and corresponds to the analytic continuation of the cross-ratio $u_1$ (cf.~\eqn{eq:MRK_limit}) by
\begin{equation}
	u_1\rightarrow e^{-2\pi i}u_1\,.
\end{equation}
This so-called \textit{Mandelstam region} corresponds to the configuration
\begin{equation}
	s,s_2>0\,,\quad s_1,s_3,s_{012},s_{123}<0\,.
\end{equation}
In \rcite{BFKLPomeron} it is shown that the BDS ansatz breaks down in the Mandelstam region starting at two loops because of contributions from a Regge cut due to a discontinuity in the variable $s_2$. While canceling in other kinematic regions, these contributions do not vanish in the Mandelstam region. Accordingly, a correction of the BDS ansatz is necessary.

The amplitude can be calculated in the Mandelstam region \cite{BFKLPomeron,reggecut,BFKLapproach} and one can write the amplitude $\mathcal{A}_{2\rightarrow4}$ (cf.~\eqn{eq:BDSdividedout}) in leading logarithmic approximation as a product of BDS ansatz $\mathcal{A}_{2\rightarrow4}^{\text{BDS}}$ and remainder function $R_6$, i.e.
\begin{equation}
	\label{eq:remainder6}
	\mathcal{A}_{2\rightarrow4}=\mathcal{A}_{2\rightarrow4}^{\text{BDS}}e^{R_6},
\end{equation}
where the remainder function reads \cite{Dixon:2012yy}
\begin{align}
	\label{eq:R6full}
		e^{R_6+i\pi\delta}|_{\text{MRK}}&=\cos\pi\omega_{ab}+i\frac{a}2 \sum_{n=-\infty}^\infty (-1)^n\left(\frac{w}{w^*}\right)^{\frac{n}2}\int_{-\infty}^\infty\frac{d\nu}{\nu^2+\frac{n^2}4}|w|^{2i\nu}\Phi_{\text{Reg}}(\nu,n)\notag\\
		&\hspace{20ex}\times \exp\left[-\omega(\nu,n)\left(\log(1-u_1)+i\pi+\frac12\log\frac{|w|^2}{|1+w|^4}\right)\right].
\end{align}
This is an all-loop expression written in terms of the cross-ratio $u_1$ and the anharmonic ratios $w$ and $w^*$ (cf.~\eqn{eq:crossratios}). 

\Eqn{eq:R6full} contains two parts originating in the BFKL equation: one part coming from a so-called Regge pole ($\cos\pi\omega_{ab}$) and a part coming from a Regge cut (sum and integral). The first part is simple, as $\omega_{ab}$ is directly related to the cusp anomalous dimension, which is known to all orders, as described below. The second part has more structure: the Regge cut is closely tied to the BFKL equation, which explains the appearance of the BFKL eigenvalue $\omega(\nu,n)$ and of the Mellin transformation integral. This second part will be of central interest for us below.

\medskip
\noindent Let us list and comment on the ingredients of \eqn{eq:R6full} individually:
\begin{itemize}
\item $a$ is the loop-counting parameter defined in \eqn{eqn:BDSloopcounting}.
\item the term $+i\pi\delta$ in the exponent on the left side of the equation cancels a divergence of the BDS ansatz and the term $\cos\pi\omega_{ab}$ comes from the Regge pole contribution, with
\begin{equation}
	\label{eq:delta_omega}
	\delta=\frac18\gamma_K(a)\log\frac{|w|^2}{|1+w|^4}\,,\quad \omega_{ab}=\frac18\gamma_K(a)\log|w|^2
\end{equation}
and the cusp anomalous dimension
\begin{equation}
	\label{eq:cuspdim}
	\gamma_K(a)=4a-4\zm_2a^2+22\zm_4a^3-\left(\frac{219}2\zm_6+4\zm_3^2\right)a^4+\mathcal{O}(a^5)\,,
\end{equation}
which is known to all loops \cite{Beisert:2006ez}. 
\item the BFKL eigenvalue $\omega(\nu,n)$ can be expanded in the coupling $a$ as
\begin{equation}
	\label{eq:bfkl_eigen}
	\omega(\nu,n)=-a\left(E_{\nu,n}^{(0)}+aE_{\nu,n}^{(1)}+a^2E_{\nu,n}^{(2)}+\mathcal{O}(a^3)\right),
\end{equation}
with the lowest order BFKL eigenvalue and its first two corrections
\begin{subequations}
\begin{align}
	\label{eq:enun}
	E_{\nu,n}:=E_{\nu,n}^{(0)}&=-\frac12\frac{|n|}{\nu^2+\frac{n^2}4}+\psi\left(1+i\nu+\frac{|n|}{2}\right)+\psi\left(1-i\nu+\frac{|n|}{2}\right)+2\gamma_E\,,\\
	\label{eq:e1}
	E_{\nu,n}^{(1)}&=-\frac14D_\nu^2E_{\nu,n}+\frac12VD_\nu E_{\nu,n}-\zm_2E_{\nu,n}-3\zm_3\,,\\
	E_{\nu,n}^{(2)}&=-\frac18\Bigg(\frac16 D_\nu^4E_{\nu,n}-VD_\nu^3E_{\nu,n}+(V^2{+}2\zm_2)D_\nu^2E_{\nu,n}-V(N^2{+}8\zm_2)D_\nu E_{\nu,n}\notag\\
	\label{eq:e2}
		       &\hspace{10ex}+\zm_3(4V^2{+}N^2)+44\zm_4E_{\nu,n}+16\zm_2\zm_3+80\zm_5\Bigg)\,.
\end{align}
\end{subequations}
Here the definitions
\begin{equation}
	\label{eq:VND}
	N=\frac{n}{\nu^2+\frac{n^2}{4}}\,,\quad V=\frac{i\nu}{\nu^2+\frac{n^2}{4}}\,,\quad D_\nu=-i\partial_\nu\,,
\end{equation}
have been used and
\begin{equation}
\label{eq:polygamma}
	\psi(z)=\frac{d}{dz}\log\Gamma(z), \quad \psi^{(k)}(z)=\frac{d^k}{dz^k}\psi(z)
\end{equation}
are the polygamma functions.
\item 
the impact factor $\Phi_{\text{Reg}}$ in \eqn{eq:R6full} can be expanded in the coupling as
\begin{equation}
	\Phi_{\text{Reg}}(\nu,n)=1+a\Phi_{\text{Reg}}^{(1)}(\nu,n)+a^2\Phi_{\text{Reg}}^{(2)}(\nu,n)+\mathcal{O}(a^3)\,,
\end{equation}
where the first two expansion factors read
\begin{subequations}
\begin{align}
	\label{eq:phi1}
	\Phi_{\text{Reg}}^{(1)}(\nu,n)&=-\frac12E_{\nu,n}^2-\frac38\frac{n^2}{(\nu^2+\frac{n^2}4)^2}-\zm_2\,,\\
		\Phi_{\text{Reg}}^{(2)}(\nu,n)&=\frac12\left(\Phi_{\text{Reg}}^{(1)}\right)^2-E_{\nu,n}^{(1)}E_{\nu,n}+\frac18\left(D_\nu E_{\nu,n}\right)^2+\frac{5}{64}N^2\left(N^2+4V^2\right)\notag\\
	\label{eq:phi2}
		&\quad-\frac{\zm_2}4\left(2(E_{\nu,n})^2+N^2+6V^2\right)+\frac{17\zm_4}4\,.
\end{align}
\end{subequations}
\end{itemize}

\subsection{Core integral for the six-point remainder function}
\label{sec:shorthand}
Evaluating \eqn{eq:R6full}, the remainder function $R_6$ can be computed order by order in $a$ (loop counter) and in $\log(1-u_1)$ after expanding both the impact factor $\Phi_{\text{Reg}}$ and the BFKL eigenvalue $\omega(\nu,n)$. 
In order to simplify the notation in such calculations, the authors of ref.~\cite{Dixon:2012yy} introduced the integral shorthand $\CI_6[\CF]$ for an integrand $\CF$, which -- due to its omnipresence in the following derivations -- will be referred to as \textit{core functional} of the six-point remainder function. The core functional is defined by
\begin{equation}
	\label{eqn:R6shorthand}
	\CI_6[\CF(\nu,n)]=\frac1{\pi}\sum_{n=-\infty}^\infty (-1)^n\left(\frac{w}{w^*}\right)^{\frac{n}2}\int_{-\infty}^\infty\frac{d\nu}{\nu^2+\frac{n^2}4} |w|^{2i\nu}\CF(\nu,n)\,.
\end{equation}
The properties of this integral have been content of various publications in which solutions have been obtained to very high loop orders and logarithmic orders \cite{BroedelSprenger,DelDuca:2019tur,Dixon:2012yy}. 
As the core functional will be the main player in \secref{sec:doublecopy} below, let us get some familiarity with its properties. 

An important property of the functions $\CF$, is that only a limited variety of terms can appear. In fact, at each loop order one finds a polynomial of the variables $E_{\nu,n}$, $V$, $N$ and $D_\nu$, defined in \eqns{eq:enun}{eq:VND}. For the first few orders one can already see that this is true from the terms provided in eqns.~\eqref{eq:e1}, \eqref{eq:e2}, \eqref{eq:phi1} and \eqref{eq:phi2}. In fact, it is sufficient to exclusively consider functions where the derivative $D_\nu$ acts on $E_{\nu,n}$, because its action on $V$ and $N$ can be rewritten in terms of combinations of $V$ and $N$:
\begin{equation}
	\label{eq:derid}
	D_\nu V=\frac{N^2}4+V^2,\quad D_\nu N=2NV.
\end{equation}
On top of the constraints concerning the constituents, there is various symmetry arguments for $R_6$, some of which carry over to $\CF$ where they will be expressed in terms of symmetry conditions with regard to the variables $\nu$ and $n$ \cite{Dixon:2012yy}. 

First, the remainder function is invariant under 
\begin{equation}
     \Delta\rightarrow-\Delta\,,
\end{equation}
where $\Delta=(1-u_1-u_2-u_3)^2-4u_1u_2u_3$ and the cross-ratios $u_i$ have been defined at the beginning of \secref{sec:BFKLMRK}. The above symmetry implies the complex conjugation symmetry of $\CF$ under
\begin{equation}
	\label{eq:conjsym}
	w\leftrightarrow w^*\,.
\end{equation}
Second, because $R_6$ is a totally symmetric function of the cross-ratios $u_1$, $u_2$ and $u_3$, it has an inversion symmetry 
\begin{equation}
	\label{eq:invsym}
	(w,w^*)\leftrightarrow (1/w,1/w^*)\,,
\end{equation}
since this transformation corresponds to $\tilde{u}_2\leftrightarrow\tilde{u}_3$ (cf.~\eqn{eq:crossratios}).

Accordingly, all core functionals should be subject to these two symmetries. Expressed in terms of the variables $\nu$ and $n$ from \eqn{eqn:R6shorthand}, \eqns{eq:conjsym}{eq:invsym} translate into
\begin{equation}
	\label{eq:paritytransforms}
	[n\leftrightarrow-n] \text{ and } [\nu\leftrightarrow-\nu,n\leftrightarrow-n]\,,
\end{equation}
respectively. Invariance under \eqns{eq:conjsym}{eq:invsym} confines $\CF(\nu,n)$ to the sector with eigenvalues $[+,+]$ exclusively.  

How to choose constituents $N$, $V$, $D_\nu$ and $E_{\nu,n}$ to yield the correct eigenvalues for the integrand $\CF(\nu,n)$? The properties of $N$, $V$, $D_\nu$ and $E_{\nu,n}$ under the symmetries are shown in table \ref{tab:parity}.
\begin{table}[t]
	\centering
	\begin{tabular}{c|c}
		& $[n\leftrightarrow-n]$, $[\nu\leftrightarrow-\nu,n\leftrightarrow-n]$ \\ \hline
		$D_\nu$  & $[+,-]$                                  \\
		$V$      & $[+,-]$                                  \\
		$N$      & $[-,-]$                                  \\
		$E_{\nu,n}$ & $[+,+]$                                 
	\end{tabular}
	\caption{Parity of the terms $D_\nu$, $V$, $N$ and $E_{\nu,n}$ under the $\mathbb{Z}_2\times\mathbb{Z}_2$ transformations $[n\leftrightarrow-n]$ and $[\nu\leftrightarrow-\nu,n\leftrightarrow-n]$, as described in ref.~\cite{Dixon:2012yy}.}
	\label{tab:parity}
\end{table}
Accordingly, the most straightforward constraint is that $\CF(\nu,n)$ should possess even powers of $N$ only. Furthermore, the powers of $D_\nu$ and $V$ in $\CF(\nu,n)$ must add up to an even number. 

Dixon, Duhr and Pennington conjectured in ref.~\cite{Dixon:2012yy} that at each order in $a$ the remainder function is given through combinations of zeta values and single-valued polylogarithms\footnote{This is conjectured because in their calculation only terms of the kind $H_{s_1}(-w)H_{s_2}(-w^*)$ appear (with $H_s(z)$ denoting harmonic polylogarithms \cite{Remiddi:2000}), which are the basic terms that give single-valued polylogarithms.} (see \secref{sec:polylogszeta} and \appref{app:conSVHPL}) in the variables $(z,\overline{z})=(-w,-w^*)$. These combinations have uniform transcendental weight for each loop order and the $[+,+]$-symmetry of the $\mathbb{Z}_2\times\mathbb{Z}_2$ transformation corresponds to conjugation and inversion in $(z,\zb)$ space as described for $(w,w^*)$ above.

While the approach for the remainder function from the BFKL method that were presented in this section gave exact values for the BFKL eigenvalue and the impact factor for the first few loop orders, calculations for the higher-loop corrections get more involved. Another way to calculate the MHV remainder function and its constituents is through Wilson loops using the AdS/CFT correspondence \cite{Alday:2007, Brandhuber:2008, Brandhuber:2010}. This method gives an all-order expression for the BFKL eigenvalue \cite{Basso:2015} and thus a better structural understanding of the MHV remainder function also at higher-loop order.


\section{Double-copy relation for the six-point remainder function}
\label{sec:doublecopy}

In this section, we will derive a double-copy relation for the six-point remainder function in planar \Nfour sYM theory in MRK. The manipulations of the core functional $\CI_6$ in \eqn{eqn:R6shorthand} are done in two steps:
\begin{enumerate}[leftmargin=1.53cm, itemindent=0cm]
	\item[\textbf{Step 1}:] Cauchy's theorem is used to rewrite the infinite sum in the core formula \eqn{eqn:R6shorthand} into a line integral in the complex plane. This can be done after identifying a suitable function whose collection of residues will comprise all terms in the sum over $n$. This will leave us with a double integral expression for the core functional. 
	\item[\textbf{Step 2}:] The two integrations are disentangled along the lines of the KLT construction. Several technical steps are required and the price to pay for the separation is an outer integration implementing the role of the sine factors in the original KLT construction. The outer integration is yet another iterated integration of polylogarithmic type.\\
The remaining inner integration in the variables $\chi_\pm$ can be recognized as an inverse Mellin transformation. This results in the desired double-copy relation. 
\end{enumerate}

\noindent Schematically, the procedure can be depicted by 
\begin{equation*}
	\boxed{\hspace*{5pt}\CI_6[\ldots]\sim\sum_{n=-\infty}^\infty \int_{-\infty}^\infty d\nu\ldots\,} \stackrel{\mathrm{Step}\, 1}{\xrightarrow{\hspace*{7ex}}}\boxed{\,\,\iint d^2\chi\ldots\hspace{-3ex}\phantom{\sum_{0}^{1}}}\stackrel{\mathrm{Step}\, 2}{\xrightarrow{\hspace*{7ex}}}\boxed{\,\,\int_0^w \frac{du}{u+1}\int d\xp\ldots \int d\xm\ldots\hspace{-3ex}\phantom{\sum_{0}^{1}}}\, .
\end{equation*}

\subsection{First step: from an infinite sum to an integral}

In order to rewrite the sum in \eqn{eqn:R6shorthand} into an integral, we need to identify a function $f$ such that  
\begin{equation}
\label{eq:goal}
	\int_{-\infty}^\infty f(z,\nu,w,w^*)\,dz=\sum_{n=-\infty}^\infty (-1)^n\left(\frac{w}{w^*}\right)^{\frac{n}2}\frac{\CF(\nu,n)}{\nu^2+\frac{n^2}4}\,,
\end{equation}
by using Cauchy's theorem when closing the integration contour. This way, one can bring the general integral formula \eqn{eqn:R6shorthand} into the form of a double integral
\begin{equation}
	\CI_6[\CF(\nu,n)]=\frac1{\pi}\int_{-\infty}^\infty d\nu\int_{-\infty}^\infty dz |w|^{2i\nu} f(z,\nu,w,w^*)\,.
\end{equation}
Considered for a fixed real value of $\nu$, the function we suggest reads
\begin{equation}
\label{eq:function}
f(z,\nu,w,w^*)=-\frac{\cosh\left(\left(\zpih\right)\phi\right)}{\cosh(\pi z)}\frac{\CF\left(\nu,-i\left(\zpih\right)\right)}{\nu^2-\frac{1}{4}\left(\zpih\right)^2}\,,
\end{equation}
where we defined\footnote{Further below, $\phi$ will be rewritten in terms of the anharmonic ratios $w$ and $w^*$ again, but for now $\phi$ allows for more compact formul\ae{}.}
\begin{equation}
	\label{eqn:wwastphi}
	w=re^{i\phi},\quad\log\frac{w}{w^*}=2i\phi \quad\text{with}\quad \phi\in[-\pi,\pi)\,.
\end{equation}
The above transformation, in which an infinite alternating sum is rewritten as a complex integral is called a \textit{Sommerfeld--Watson transformation} \cite{NIST}.

\paragraph{$\CF$ as a function of $z$.} Let us check how the function $\CF$ in \eqn{eq:function} can now be written in terms of the variable $z$: as discussed in section~\ref{sec:shorthand} above, functions $\CF$ appearing inside the core functional \eqn{eqn:R6shorthand} can be expressed as combinations of the variables $E_{\nu,n}$, $V$, $N$ and $D_\nu$ defined in \eqns{eq:enun}{eq:VND}. While $V$ and $N$ are clearly meromorphic functions in $n$, we have to investigate the behavior of $|n|$ in $E_{\nu,n}$ more closely. For concise notation, we will split the eigenvalue $E_{\nu,n}$ further into 
\begin{align}
	\label{eq:E}
	E_{\nu,n}&=-\frac{|n|}{n}\frac{N}{2}+E_\psi\,,\quad\text{ where }\notag\\
	E_\psi&=\psi\left(1+\tfrac{|n|}{2}+i\nu\right)+\psi\left(1+\tfrac{|n|}{2}-i\nu\right) - 2\psi(1)\,.
\end{align}
The function $\CF$ has to exhibit $[+,+]$-symmetry, that is, it needs to have positive unit eigenvalue for either of $n\to-n$ and $\nu\to -\nu$. Accordingly, $\CF$ must contain an even power of $N$, and the powers of $D_\nu$ and $V$ must add up to an even number (cf.~\secref{sec:shorthand}).

The requirement of $[+,+]$-symmetry is advantageous here: it allows to replace $n$ by $-n$ and $\nu$ by $-\nu$, which allows us to get rid of the absolute values in the expression for $E_\psi$ in \eqn{eq:E}. In turn, after removing the absolute values, the whole function $\CF$ will be meromorphic in the upper half-plane $z$ and is thus ready for using Cauchy's theorem below.  

At the locus of the residues from \eqn{eq:function} contributing for our choice of contour, that is at $n=-i\left(\zpih\right)$, the variables $E_\psi$, $V$, $N$ read  
\begin{subequations}
	\begin{align}
		\label{eqn:NVEnew1}
		N&=\frac{-i\left(\zpih\right)}{\nu^2-\frac{1}{4}\left(\zpih\right)^2}\,,\quad V=\frac{i\nu}{\nu^2-\frac{1}{4}\left(\zpih\right)^2}\,,\\
		\label{eqn:NVEnew2}E_\psi
		&=\psi\left(\tfrac54-i\left(\tfrac{z}{2}{-}\nu\right)\right)+\psi\left(\tfrac54-i\left(\tfrac{z}{2}{+}\nu\right)\right) - 2\psi(1)\,,
	\end{align}
\end{subequations}
whereas $D_\nu$ remains unchanged. The analytic continuation of $\CF(\nu,n)$ described here will be named $\CF_+(\nu,n)$ henceforth.

\subsubsection{Applying Cauchy's theorem} 
\begin{figure}[t]
\begin{center}
\begin{tikzpicture}[decoration={markings,
mark=at position 4.5cm with {\arrow[line width=2pt]{>}},
mark=at position 13cm with {\arrow[line width=2pt]{>}},
mark=at position 22.8cm with {\arrow[line width=2pt]{>}}
}
]
\draw[->] (-6.2,0) -- (6.2,0) coordinate (xaxis);
\draw[->] (0,-2) -- (0,6) coordinate (yaxis);

\fill (0,0.5)  circle[radius=2.5pt] node[right] {$z=\frac12i$};
\fill (0,1.5)  circle[radius=2.5pt] node[right] {$z=\frac32i$};
\fill (0,2.5)  circle[radius=2.5pt] node[right] {$z=\frac52i$};
\fill (0,3.5)  circle[radius=2.5pt] node[right] {$\quad\quad{\vdots}$};
\fill (0,-0.5)  circle[radius=2.5pt] node[right] {$z=-\frac12i$};
\fill (0,-1.5)  circle[radius=2.5pt] node[right] {$\quad\quad{\vdots}$};

\fill[blue] (-3,-0.5)  circle[radius=2.5pt] node[below] {$z=-2\nu-\frac{i}2$};
\fill[blue] (3,-0.5)  circle[radius=2.5pt] node[below] {$z=2\nu-\frac{i}2$};

\fill[red] (1.5,-2)  circle[radius=2.5pt] node[right] {$z=\nu-\frac52i$};
\fill[red] (1.5,-3)  circle[radius=2.5pt] node[right] {$z=\nu-\frac92i$};
\fill[red] (1.5,-4)  circle[radius=2.5pt] node[right] {$\quad\quad{\vdots}$};

\fill[red] (-1.5,-2)  circle[radius=2.5pt] node[left] {$z=-\nu-\frac52i$};
\fill[red] (-1.5,-3)  circle[radius=2.5pt] node[left] {$z=-\nu-\frac92i$};
\fill[red] (-1.5,-4)  circle[radius=2.5pt] node[left] {${\vdots}\quad\quad$};

\path[draw,line width=1.8pt,postaction=decorate](5.5,0) node[below] {$R$} arc (0:180:5.5)   node[below] {$-R$} -- (5.5,0);

\node[below] at (xaxis) {$\operatorname{Re} z$};
\node[left] at (yaxis) {$\operatorname{Im} z$};
\node[below left] {$0$};
\node at (4,4.5) {$\gamma_{R}$};
\end{tikzpicture}
\caption{Positions of the poles of the function eq.~(\ref{eq:function}) to be considered when rewriting the sum in eq.~(\ref{eq:goal}) as an integral. Black dots denote poles from the $\cosh(\pi z)^{-1}$ term in $f$, blue dots refer to the poles from $(\nu^2-(z+i/2)^2/4)^{-1}$. Poles marked in red can be caused by terms $E_\psi$ (and derivatives thereof) occurring in $\CF_+$.}
\label{fig:contour}
\end{center}
\end{figure}
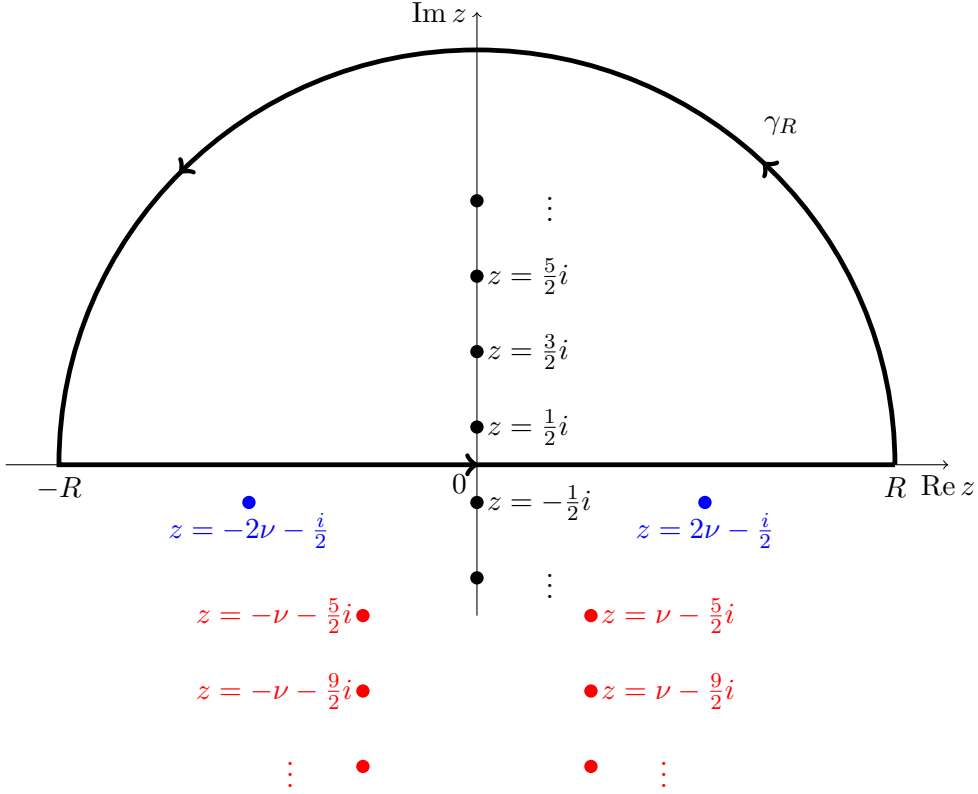

In order to get the relation \eqn{eq:goal} using the function \eqn{eq:function}, we will close the integration contour in the upper complex plane and employ Cauchy's integral formula. Our contour as well as all poles of the function $f$ are depicted in \figref{fig:contour}. The function $f$ possesses infinitely many poles on the imaginary axis (black dots) as well as two poles in the lower half-plane (blue dots). In addition, one has to consider poles due to polygamma terms in $\CF_+$ (red dots). 
Closing the contour in the upper half-plane, only poles on the positive imaginary axis will contribute to the integral. Nevertheless, we will comment on all poles of $f$ for completeness. 

\paragraph{Poles on the imaginary axis.}
\label{ssec:polesres}
We will first consider the poles of $f$ on the imaginary axis of which only the ones on the positive imaginary axis will contribute to the integral when closing the contour in the upper half-plane. Since $\cosh(i\xi)=\cos(\xi)$, the function $f$ in \eqn{eq:function} has poles for $z=i\frac{2n-1}{2}$, $n\in\mathbb{Z}$. These poles are of first order and the associated residues read
\begin{subequations}
\begin{align}
	\operatorname{Res}_{i\frac{2n-1}{2}}f&=\lim_{z\rightarrow i\frac{2n-1}{2}}\left(z-i\left(\frac{2n{-}1}{2}\right)\right)\left(-\frac{\cosh\left(\left(\zpih\right)\phi\right)}{\cosh(\pi z)}\frac{\CF_+\left(\nu,-i\left(\zpih\right)\right)}{\nu^2-\frac{1}{4}\left(\zpih\right)^2}\right)\\
		&=\frac{(-1)^n}{2\pi i}\frac{\CF_+(\nu,n)}{\nu^2+\frac{n^2}4}\left(e^{ni\phi}+e^{-ni\phi}\right)
		=\frac{(-1)^n}{2\pi i}\frac{\CF_+(\nu,n)}{\nu^2+\frac{n^2}{4}}\left(\left(\frac{w}{w^*}\right)^{\frac{n}2}+\left(\frac{w}{w^*}\right)^{-\frac{n}2}\right).\label{eqn:nn}
\end{align}
\end{subequations}
When calculating the above residue, one has to ensure regularity of $\CF_+$ at points $z=i\frac{2n-1}{2}$. In other words, $\CF_+$ should not contribute additional poles at these points. In fact, considering the composition of $\CF_+$ as discussed in \secref{sec:shorthand}, an additional divergent contribution at $z=i\frac{2n-1}{2}$ can be excluded. 
When closing the integration contour in the upper half-plane, all residues with positive $n$ are included. For each of them, the two contributions in \eqn{eqn:nn} deliver all terms with positive and negative $n$ in the sum in \eqn{eq:goal}. The only contribution in the sum not covered by the contour integral is the one for $n=0$: this will have to be taken care of later explicitly. 

\paragraph{Two poles from denominators (lower half-plane).}
\label{sec:2poles}
There are two additional poles of $f$ at $z=\pm2\nu-\frac{i}2$ originating from the term $\nu^2-\frac{1}{4}\left(\zpih\right)^2$ in the denominator in $f$ and in the denominators of possible $V$ and $N$ terms in $\CF_+$. While these are simple poles for $\CF_+=1$, their order increases with every insertion of $N$, $V$ or $D_\nu$ in $\CF_+$. Located in the lower half-plane, they do not contribute for our choice of contour. 

\paragraph{Poles from polygamma terms in $\CF_+$ (lower half-plane).}
\label{ssec:possible_terms}
Lastly, using the expression in \eqn{eqn:NVEnew2} we need to check for potential further poles coming from terms of $E_\psi$ and its $D_\nu$-derivatives in the function~$\CF_+$. Thus, we consider the singularities of the polygamma functions $\psi^{(k)}(\zeta)$ defined in \eqn{eq:polygamma}, which have poles only for $\zeta\,{=}-m$ with $m\in\mathbb{N}_0$. In \eqn{eqn:NVEnew2}, this pole condition translates into $z=\mp\nu-i(2m+5/2)$, $m\in\mathbb{N}_0$. For real $\nu$, those poles will again reside in the lower half-plane, so there is no problem when closing the contour above the real axis.

\bigskip
In summary, both the poles from the denominators of $f$ as well as the poles from polygamma terms considered in this subsection do not contribute for our choice of contour, which proves $f$ to indeed be a suitable function for converting the sum in \eqn{eqn:R6shorthand} into an integral. 

\subsubsection{Double integral}
\label{sec:doubleint}
Closing the contour as depicted in \figref{fig:contour} amounts to evaluating 
\begin{equation}
	\label{eq:412firsteqn}
	\lim_{R\rightarrow\infty}\left(\int_{-R}^R \!f(z)dz+\int_{\gamma_R}\!\!f(z)dz\right)=2\pi i\sum_{n\in\mathbb{N}}\operatorname{Res}_{i\left(\frac{2n-1}{2}\right)}f=\sum_{n\in\mathbb{Z}\backslash\{0\}}\!\!(-1)^n\left(\frac{w}{w^*}\right)^{\frac{n}2}\frac{\CF(\nu,n)}{\nu^2+\frac{n^2}2}\,,
\end{equation}
where the semi-circle is parameterized by
\begin{equation}
\label{eq:contour}
	\gamma_R(t)=Re^{it}\,,\quad t\in[0,\pi]\,,\quad R\in\mathbb{N}\,,\quad R\gg1\,. 
\end{equation}
Choosing an integer radius $R$ avoids the contour hitting any of the poles in the upper half-plane. In order to get the desired relation \eqn{eq:goal}, one needs to show that $\int_{\gamma_R} f(z)dz$ vanishes for $R\rightarrow\infty$. This is indeed the case, as is proven in appendix \ref{app:proof1}.  

Taking finally the ($n=0$)-contribution into account, the core functional of the remainder function can now indeed be written as the double integral
\begin{align}
\label{eq:double_int}
\CI_6\left[\CF_+\left(\nu,-i\left(\zpih\right)\right)\right]=&-\frac1{\pi}\int_{-\infty}^\infty d\nu\int_{-\infty}^\infty dz |w|^{2i\nu}\frac{\cosh\left(\left(\zpih\right)\phi\right)}{\cosh(\pi z)}\frac{\CF_+\left(\nu,-i\left(\zpih\right)\right)}{\nu^2-\frac{1}{4}\left(\zpih\right)^2}\nonumber\\
	&+\frac{1}{\pi}\int_{-\infty}^\infty d\nu |w|^{2i\nu}\frac{\CF_+(\nu,0)}{\nu^2}\,.
\end{align}
Restoring the variables $w, w^\ast$ via \eqn{eqn:wwastphi} and using the identity $\frac{\pi}{\cosh(\pi x)}=\Gammafn\left(\tfrac12{+}ix\right)\Gammafn\left(\tfrac12{-}ix\right)$ the result can be rewritten as
\begin{align}
	\label{eq:I(nuz)2}
	&\CI_6\left[\CF_+\left(\nu,-i\left(\zpih\right)\right)\right]=\nonumber\\
	&\hspace{4ex}-\frac{1}{2\pi^2}\int_{-\infty}^\infty \!\!\!d\nu\int_{-\infty}^\infty dz\Gammafn\left(\tfrac12{+}iz\right)
	\Gammafn\left(\tfrac12{-}iz\right) w^{i\left(\nu-\frac{z}2\right)+\frac14}(w^*)^{i\left(\nu+\frac{z}2\right)-\frac14}\frac{\CF_+\left(\nu,-i\left(\zpih\right)\right)}{\nu^2-\frac{1}{4}\left(\zpih\right)^2}\nonumber\\
	&\hspace{4ex}+(w\leftrightarrow w^*)+\frac{1}{\pi}\int_{-\infty}^\infty d\nu |w|^{2i\nu}\frac{\CF_+(\nu,0)}{\nu^2}\,,
\end{align}
where the integral for the ($n=0$)-term is to be understood as a principal value integral: around the pole on the real axis a small $\ve$-neighborhood is excluded for integration, where $\ve$ is taken to zero after integration. This completes step~1.

\subsection{Second step: separating the integrations}
\label{sec:separating}
When considering the integral representation of $\CI_6$ in \eqn{eq:I(nuz)2} above, the main obstacle disentangling the two integrations in the way KLT did for the closed string integral is the fraction term $(\nu^2-(z+i/2)^2/4)^{-1}$. Performing a change of variables from $(\nu,z)$ to the new variables\footnote{The variables in \eqn{eqn:changeofvariables} have already been employed successfully in \cite{BroedelSprenger}.}
\begin{equation}
	\label{eqn:changeofvariables}
	\chi_\pm=-i\left(\frac{z}2\pm\nu\right)+\frac14\,,
\end{equation}
one finds 
\begin{alignat}{2}
	N=&\frac1{\xp}+\frac1{\xm}\,,\quad &V&=\frac1{2\xp}-\frac1{2\xm}\,,\notag\\ 
	\label{eqn:newNVDE}
	D_\nu=&-\partial_{\xp}+\partial_{\xm}\,,
	\quad &E_\psi&=\psi(1{+}\xp)+\psi(1{+}\xm)+2\gamma_E\,.
\end{alignat}
Given the analytical structure of $\CF_+$, that is, the particular appearance of the terms in \eqn{eqn:newNVDE}, it is now possible to rewrite
\begin{equation}
	\label{eq:Fseparated}
	\CF_+[\vec{\chi}]=\sum_{i}f_i(\xp)g_i(\xm),
\end{equation}
where $f_i$ and $g_i$ are a finite set of functions and $\CF_+[\vec{\chi}]=\CF_+\left[\frac{i}2(\xp{-}\xm),\xp{+}\xm\right]$. Several examples for the above rewriting are collected in \appref{app:calculations}. 

In the new set of integration variables $\chi_\pm$ the core functional reads
\begin{align}
\label{eq:chi}
\CI_6[\CF_+[\vec{\chi}]]&=\frac1{2\pi^2}\int_{\frac14-i\infty}^{\frac14+i\infty}d\xp\int_{\frac14-i\infty}^{\frac14+i\infty}d\xm(w^*)^{-\xp}w^{\xm}\frac{\CF_+[\vec{\chi}]}{\xp\xm}\Gammafn(1{-}\xp{-}\xm)\Gammafn(\xp{+}\xm)\nonumber\\
	&\quad+(w\leftrightarrow w^*)+\frac{1}{\pi}\int_{-\infty}^\infty d\nu |w|^{2i\nu}\frac{\CF_+(\nu,0)}{\nu^2}\,.
\end{align}
The two integrations in the first line of \eqn{eq:chi} can be identified\footnote{Our convention for the Mellin transform of a function $f:[0,\infty)\rightarrow\mathbb{C}$ (assuming it to be well-defined) reads
\begin{equation*}
	F(s)=\CM[f](s):=\int_0^\infty f(x)x^{s-1}\,dx\,,
\end{equation*} 
whereas the inverse Mellin transform is ($c\in\mathbb{R}$ such that $F$ can be integrated in $t$ on the line $c+it$)
\begin{equation*}
	f(x)=\CM^{-1}[F](x):=\frac{1}{2\pi i}\int_{c-i\infty}^{c+i\infty}x^{-s}F(s)\,ds\,.
\end{equation*}
}
as \textit{inverse Mellin transforms}, where the factors $(w^*)^{-\xp}$ and $w^{\xm}$ are the integration kernels for inverse Mellin transforms in the variables $w^*$ and $1/w$, respectively. 

The only terms preventing us from separating the two inverse Mellin transforms are the two factors with Gamma functions, which we are going to tackle in the following. 
Writing the Gamma factors in their integral representation and using the substitutions $u=t/s$ and $v=s$ yields \footnote{Starting from 
\[
	\Gammafn(1{-}\xp{-}\xm)\Gammafn(\xp{+}\xm)=\frac{\pi}{\sin(\pi(\chi_++\chi_-))}=\frac{\pi}{\sin(\pi\chi_+)\cos(\pi\chi_-)+\cos(\pi\chi_+)\sin(\pi\chi_-)}\,,
\]
one could think of cleverly expanding the last expression. However, the expansion would be valid only for a certain ranges of $\chi_\pm$, which would not allow disentangling the integrals later. }
\begin{align}
\label{eq:gamma_id}
	\Gammafn(1{-}\xp{-}\xm)\Gammafn(\xp{+}\xm)\nonumber
	&=\int_0^\infty ds\,e^{-s}s^{-\xp-\xm}\int_0^\infty\frac{dt}{t}e^{-t}\,t^{\xp+\xm}\\
	&=\int_0^\infty\int_0^\infty \frac{dt}tds\,e^{-s-t}\left(\frac{t}{s}\right)^{\xp+\xm}\nonumber\\
	&=\int_0^\infty\int_0^\infty du\,dv \frac{v}{uv}e^{-v-vu}\,u^{\xp+\xm}\nonumber\\
	&=\int_0^\infty\frac{du}uu^{\xp}u^{\xm}\underbrace{\int_0^\infty dv\,e^{-v(1+u)}}_{=\frac1{u+1}}\nonumber\\
	&=\int_0^\infty\frac{du}{u(u+1)}u^{\xp}u^{\xm}.
\end{align}%
\begin{figure}[t]
\begin{center}
\begin{tikzpicture}[decoration={markings,
mark=at position 2.7cm with {\arrow[line width=1.5pt]{>}},
mark=at position 6.5cm with {\arrow[line width=1.5pt]{>}},
mark=at position 9.8cm with {\arrow[line width=1.5pt]{>}},
mark=at position 24.8cm with {\arrow[line width=1.5pt]{>}}
}
]
\draw[->, thick] (-1.2,0) -- (6.2,0) coordinate (xaxis);
\draw[->, thick] (0,-2.5) -- (0,1.2) coordinate (yaxis);

\draw[blue, thick] (2,0) arc (0:-26.5:2);
\node[blue] at (1.6,-0.4) {$-\phi$};

\path[red,draw,line width=1pt,postaction=decorate](0.5,0) -- (5.6,0) arc (0:-26.5:5.6) -- (0.447,-0.223) arc (-26.5:0:0.5) -- (0.6,0);

\fill (2,1)  circle[radius=2.5pt] node[left] {$w$};
\fill (2,-1)  circle[radius=2.5pt] node[below] {$w^*$};

\node[below] at (xaxis) {$\operatorname{Re} u$};
\node[left] at (yaxis) {$\operatorname{Im} u$};
\node[below left] {$0$};
\node[red] at (2.7,0.3) {$\gamma_{1}$};
\node[red] at (5.8,-1.6) {$\gamma_{R}$};
\node[red] at (3.2,-2) {$\gamma_{2}$};
\node[red] at (0.25,-0.2) {$\gamma_{\varepsilon}$};
\node[red] at (5.6,0.3) {$R$};
\node[red] at (0.5,0.2) {$\varepsilon$};
\end{tikzpicture}
\caption{Rotation of the integration of the Gamma factor $\Gammafn(1-\xp-\xm)\Gammafn(\xp+\xm)$ in the complex plane. In the limit $R\rightarrow\infty$, $\varepsilon\rightarrow0$ the path $\gamma_1$ corresponds to the integration on the positive real axis, while $\gamma_2$ is the rotated integration by the angle $-\phi=-\arg(w)$ which will help to cancel the phases of $w$ and $w^*$ in the core functional. Last, the terms coming from $\gamma_R$ and $\gamma_\varepsilon$ vanish in the limits $R\rightarrow\infty$ and $\varepsilon\rightarrow0$, respectively, for fixed $\chi_\pm$ as shown in appendix~\ref{app:proof2}.}
\label{fig:gamma_id}
\end{center}
\end{figure}
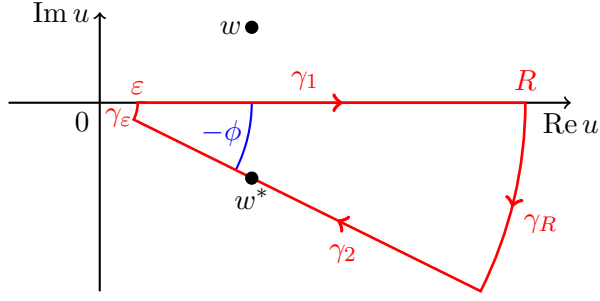%
We would like to insert this identity into \eqn{eq:chi} and then pull the $u$-integration to the front as this would disentangle the variables $\xp$ and $\xm$. But since $u/w^*$ and $uw$ have imaginary parts in \eqn{eq:chi}, the terms $(u/w^*)^{\xp}$ and $(uw)^{\xm}$ could diverge at the $\chi_\pm$-boundaries $\frac14\pm i\infty$. 

The way to proceed here is to rotate the integration path of the Gamma function identity \eqn{eq:gamma_id} by the angle $(-\arg(w))$ as depicted in \figref{fig:gamma_id} in order to get rid of these divergences. We combine the integration path in \eqn{eq:gamma_id} (i.e.~the positive real axis given by path $\gamma_1$ in \figref{fig:gamma_id}) with the rotated integration path (given by $\gamma_2$ in \figref{fig:gamma_id}) using the contours $\gamma_R$ and $\gamma_\varepsilon$ in \figref{fig:gamma_id} and send $R\rightarrow\infty$, $\varepsilon\rightarrow0$ afterwards. The contributions from the paths $\gamma_R$ and $\gamma_\varepsilon$ vanish in this limit (see appendix~\ref{app:proof2} for the proof), which renders the rotation a valid transformation. The path $\gamma_\varepsilon$ is necessary to avoid the branch point of the integrand at $u=0$.

Employing furthermore $w=re^{i\phi}$ one finds the identity
\begin{align}
\label{eq:gamma_id_shifted}
\Gammafn(1{-}\xp{-}\xm)\Gammafn(\xp{+}\xm)&=\int_0^{e^{-i\phi} \infty}\frac{du}{u(u+1)}u^{\xp}u^{\xm}\nonumber\\
	&=\int_0^\infty \frac{dt}{t(e^{-i\phi}t+1)}(e^{-i\phi}t)^{\xp}(e^{-i\phi}t)^{\xm},
\end{align}
using the substitution $t=ue^{i\phi}$. These phases cancel the phases of $w$ and $w^*$ in the $(w^*)^{-\xp}$ and $(1/w)^{-\xm}$ terms of the core functional, so that there are no divergences at the $\chi_\pm$-boundaries when inserting the identity \eqn{eq:gamma_id_shifted}. Thus, we can insert this identity into \eqn{eq:chi} and pull the $u$-integration to the front. This yields, using $\CF_+=\sum_if_i(\xp)g_i(\xm)$ (cf.~\eqn{eq:Fseparated})
\begin{align}
	\CI_6[\CF_+]=&\sum_i\frac1{2\pi^2}\int_0^\infty\frac{dt}{t\left(te^{-i\varphi}+1\right)}\int_{\frac14-i\infty}^{\frac14+i\infty}\frac{d\xp}{\xp}\left(\frac{r}{t}\right)^{-\xp}f_i(\xp) \int_{\frac14-i\infty}^{\frac14+i\infty}\frac{d\xm}{\xm}\left(\frac1{t\,r}\right)^{-\xm}g_i(\xm)\notag\\
	&+(w\leftrightarrow w^*)+\frac{1}{\pi}\int_{-\infty}^\infty d\nu |w|^{2i\nu}\frac{\CF_+(\nu,0)}{\nu^2}\,.
\end{align}
As the next transformation, we substitute $s=1/t$ to find ($w=re^{i\phi}$, cf.~\eqn{eqn:wwastphi}) 
\begin{align}
	\CI_6[\CF_+]=&\sum_i\frac1{2\pi^2}\int_0^\infty\frac{ds}{s+e^{-i\varphi}}\int_{\frac14-i\infty}^{\frac14+i\infty}\frac{d\xp}{\xp}(rs)^{-\xp}f_i(\xp) \int_{\frac14-i\infty}^{\frac14+i\infty}\frac{d\xm}{\xm}\left(\frac{s}{r}\right)^{-\xm}g_i(\xm)\nonumber\\
	&+(w\leftrightarrow w^*)+\frac{1}{\pi}\int_{-\infty}^\infty d\nu |w|^{2i\nu}\frac{\CF_+(\nu,0)}{\nu^2}\nonumber\\
	=&-2\sum_i\int_0^\infty\frac{ds}{s+e^{-i\varphi}}\CM^{-1}\left[\frac{f_i(\xp)}{\xp}\right](rs)\CM^{-1}\left[\frac{g_i(\xm)}{\xm}\right]\left(\frac{s}r\right)\notag\\
        \label{eq:full_s}
	&+(w\leftrightarrow w^*)+\frac{1}{\pi}\int_{-\infty}^\infty d\nu |w|^{2i\nu}\frac{\CF_+(\nu,0)}{\nu^2}\,.
\end{align}%
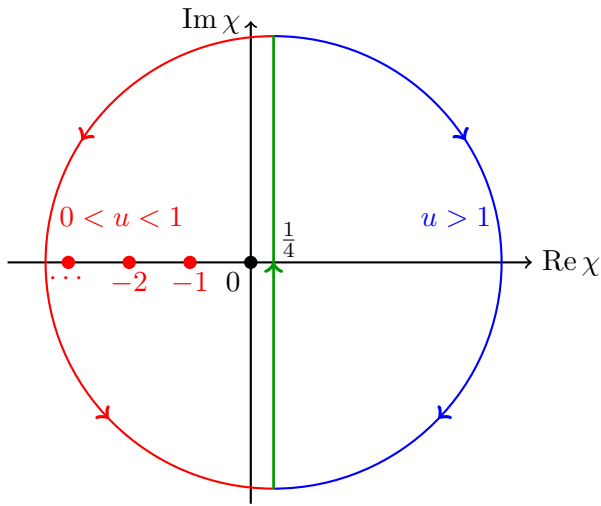
\begin{figure}[t]
\begin{center}
\begin{tikzpicture}[decoration={markings,
mark=at position 3cm with {\arrow[line width=1.5pt]{>}},
mark=at position 7cm with {\arrow[line width=1.5pt]{>}},
mark=at position 10cm with {\arrow[line width=1.5pt]{>}},
mark=at position 24.8cm with {\arrow[line width=1.5pt]{>}}
}
]
\draw[->, thick] (-3.2,0) -- (3.7,0) coordinate (xaxis);
\draw[->, thick] (0,-3.2) -- (0,3.2) coordinate (yaxis);

\draw[blue, thick, postaction=decorate] (0.3,3) arc (90:-90:3);
\node[blue] at (2.7,+0.6) {$u>1$};
\draw[red, thick, postaction=decorate] (0.3,3) arc (90:270:3);
\node[red] at (-1.7,0.6) {$0<u<1$};

\path[draw,green!60!black,line width=1pt,postaction=decorate](0.3,-3) -- (0.3,3);

\node[right] at (xaxis) {$\operatorname{Re} \chi$};
\node[left] at (yaxis) {$\operatorname{Im} \chi$};
\node[below left] {$0$};
\node at (0.5,0.3) {$\frac14$};

\fill (0,0)  circle[radius=2.5pt];
\fill[red] (-0.8,0)  circle[radius=2.5pt] node[below]{$-1$};
\fill[red] (-1.6,0)  circle[radius=2.5pt] node[below]{$-2$};
\fill[red] (-2.4,0)  circle[radius=2.5pt] node[below]{$\cdots$};
\end{tikzpicture}
\caption{Integration contour of the inverse Mellin transform $\CM^{-1}\left[f(\chi)/\chi\right](u)$. In our case the integration is parallel to the imaginary axis with a real shift of $1/4$ (green line). For $u>1$ we can close the contour on the right side (blue curve),
	and for $0<u<1$ we close the contour on the left side (red path), collecting the residue at $\chi=0$ (black dot) as well as possible residues at $\chi=-n$, $n\in\mathbb{N}$, originating from polygamma terms (red dots), in the case $0<u<1$. 
	}
\label{fig:mellin_contour}
\end{center}
\end{figure}%
We want to see how the inverse Mellin transformations behave. For this we assume $r\,{=}\,|w|\,{<}\,1$, where for the case $r>1$ an analogous calculation is possible. For an inverse Mellin transform $\CM^{-1}[f(\chi)/\chi](u)$, $u\in\mathbb{R}_+$, we close the integration contour (which is parallel to the imaginary axis) on the right side of the complex plane for $u>1$ and on the left side for $0<u<1$, as shown in \figref{fig:mellin_contour}. Thus, we need to look at the poles of the function $f(\chi)/\chi$ in the inverse Mellin transform. This function will be of the general form (compare with basis functions of ref.~\cite{BroedelSprenger}) 
\begin{equation}
	f_{k,m,\{a_j\}}(\chi)=\frac{1}{\chi^k}\left[\prod_j\psi^{(a_j)}(1+\chi)\right](\psi(1+\chi)+\gamma_E)^m,
\end{equation}
with $k,m, a_j\in\mathbb{N}_0$ and $\psi^{(\ell)}$ the polygamma functions from \eqn{eq:polygamma}\footnote{In order to avoid cluttered notation, the subindices of the function $f$ will be collected in a common index $i$.}. This function has a pole at $\chi=0$ and poles at $-\chi\in\mathbb{N}$, if $f$ contains polygamma terms. All these poles are located to the left of our integration contour (since the contour is parallel to the real axis with a real shift of $+1/4$). Thus, they only contribute for the case $0<u<1$, so that the inverse Mellin transform $\CM^{-1}[f(\chi)/\chi](u)$ evaluates to zero if $u>1$. 

With this insight we can simplify the $s$-integration in \eqn{eq:full_s}. Namely, in this equation we then need both $rs<1$ as well as $s/r<1$ in order to have a non-zero contribution to the integral. Since we assumed $r<1$, these two inequalities are only both fulfilled for $s<r$. This means that we can use $r$ as the upper limit for the $s$-integration in \eqn{eq:full_s}. Last, we substitute $u=se^{i\phi}$ and arrive at the final formula
\begin{align}
	\CI_6[\CF_+]=&-2\sum_i\int_0^w\frac{du}{u+1}\CM^{-1}\left[\frac{f_i(\xp)}{\xp}\right](uw^*)\CM^{-1}\left[\frac{g_i(\xm)}{\xm}\right]\left(\frac{u}{w}\right)\notag\\
\label{eq:newresult}
		    &+(w\leftrightarrow w^*)+\frac{1}{\pi}\int_{-\infty}^\infty d\nu |w|^{2i\nu}\frac{\CF_+(\nu,0)}{\nu^2}\,.
\end{align}
\noindent Examples for the calculations of $\CI_6[1]$, $\CI_6[V]$, $\CI_6[N]$ and $\CI_6[E_\psi]$ using this formula are shown in appendix \ref{app:calculations}.

\Eqn{eq:newresult} is the final result of our transformation, and the properties of its double-copy structure can be described as follows:
\begin{itemize}
	\item The terms for $n\neq0$ from the infinite sum of the usual formulation of the remainder function turn into double-copies of two inverse Mellin transforms. This is very similar to the KLT relations that we have seen in section \ref{sec:klt}! Here, the two inverse Mellin transforms under the integral are the formal analogues to the open-string amplitudes from the KLT relations.
	\item The role of the sine prefactor in the KLT relations is taken here by the outer integral. Similar to the KLT relations, the two integrals of the inverse Mellin transforms combine to yield a single-valued object. 
	\item The ($n=0$)-term is a single-valued function of $w$ by itself as it depends on $|w|^2$ only. 
\end{itemize}
In distinction to the KLT relation, the operation connecting the two multi-valued objects is not a multiplication by a sine factor, but rather an integration. Two facts are important to mention about the outer $u$-integration:
\begin{itemize}
	\item The integration variable $u$ scales the kinematic variables in the arguments of the inverse Mellin transforms. That is, in the first integral, $w^*$ and $1/w$ are being scaled by the outer integral, and similarly, $w$ and $1/w^*$ in the second term of \eqn{eq:newresult}.
	\item The $u$-integration of the two inverse Mellin transforms comes with the measure \mbox{$(u+1)^{-1}$}. The differential $\tfrac{du}{u+1}$ comprises together with $\tfrac{du}{u}$ and $\tfrac{du}{u-1}$ the class of harmonic polylogarithms (cf.~\secref{sec:polylogszeta}). This underlines why the remainder function can be written in terms of polylogarithms: the inverse Mellin transforms will always yield powers of logarithms or polylogarithms (see appendix~\ref{app:calculations} for examples of particular inverse Mellin transforms) which can then easily be integrated with the measure $(u+1)^{-1}$.
\end{itemize}
%


\section{Towards a double-copy relation for the seven-point remainder function}
\label{sec:7point}
After reformulating the six-point remainder function as a double-copy in the last section, let us discuss and define the framework for decomposing the seven-point remainder function in the following. For the seven-point MHV amplitude, the deviation from the BDS ansatz is again expressed through a non-trivial remainder function $R_7$:
\begin{equation}
	A_7^{\text{MHV}}=A_7^{\text{BDS}}e^{R_7^{\text{MHV}}}.
\end{equation}
In the leading logarithmic approximation the seven-point remainder function reads \cite{7Point}
\begin{equation}
	R_7^{\text{MHV}}=1+i\pi\sum_{\ell=2}^\infty\sum_{k=0}^{\ell-1}\frac{a^\ell}{k!(\ell-1-k)!}\log^k(1-u_{1,1})\log^{\ell-1-k}(1-u_{1,2})\,\CI_7[E_{\nu,n}^kE_{\mu,m}^{\ell-k-1}]\,,
\end{equation}
with the core functional
\begin{align}
		&\CI_7[\CF(\nu,n,\mu,m)]:=\\
		&\quad\quad\sum_{n=-\infty}^\infty\sum_{m=-\infty}^\infty\int_{-\infty}^\infty \frac{d\nu}{2\pi}\int_{-\infty}^\infty \frac{d\mu}{2\pi}w_1^{i\nu+\frac{n}2}(w_1^*)^{i\nu-\frac{n}2}w_2^{i\mu+\frac{m}2}(w_2^*)^{i\mu-\frac{m}2} C(\nu,n,\mu,m)\,\CF(\nu,n,\mu,m).
\end{align}
The new object appearing in $\CI_7$ is the central emission vertex \cite{Bartels:2011ge}
\begin{equation}
	C(\nu,n,\mu,m)=(-1)^{n+m}\frac{\Gammafn\left(-i\nu-\frac{n}2\right)\Gammafn\left(i\mu+\frac{m}2\right)\Gammafn\left(i(\nu-\mu)+\frac{m-n}2\right)}{\Gammafn\left(1+i\nu-\frac{n}2\right)\Gammafn\left(1-i\mu+\frac{m}2\right)\Gammafn\left(1-i(\nu-\mu)+\frac{m-n}2\right)}\,,
\end{equation}
which is very reminiscent to the four-point closed-string scattering amplitude at tree level\footnote{In fact, the formal similarity between central emission vertex and the Virasoro--Shapiro amplitude triggered the starting question for the current article.}. Instead of variables $\nu$ and $n$ for integration and summation for $\CI_6$ (cf.~\eqn{eqn:R6shorthand}), respectively, there are now two integration variables $\nu,\mu$ as well as two summation variables $n$ and $m$ in $\CI_7$. The four quantities are entangled not only in the seven-point analogue of the function $\CF$, but in particular by the central emission vertex. 

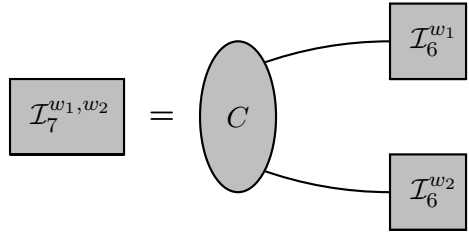
\begin{figure}[t]
	\begin{center}
				
		\begin{tikzpicture}
			
			\filldraw[thick,lightgray] (-5,-0.5) -- (-3.5,-0.5) -- (-3.5,0.5) -- (-5,0.5) -- (-5,-0.5);
			\draw[thick] (-5,-0.5) -- (-3.5,-0.5) -- (-3.5,0.5) -- (-5,0.5)-- (-5,-0.5)-- (-3.5,-0.5);
			
			\node[black] at (-4.2,0) {$\mathcal{I}_7^{w_1,w_2}$};
			
			\node[black] at (-3,0) {\textbf{$=$}};
			
			\draw[thick] (0,1) arc(90:110:5);
			\draw[thick] (0,-1) arc(270:250:5);
			
			\filldraw[thick,lightgray] (-2,0) ellipse (.5cm and 1.cm);
			\draw[thick] (-2,0) ellipse (.5cm and 1.cm);
			
			\node[black] at (-2,0) {$C$};
			
			\filldraw[thick,lightgray] (0,0.5) -- (1,0.5) -- (1,1.5) -- (0,1.5) -- (0,0.5);
			\draw[thick] (0,0.5) -- (1,0.5) -- (1,1.5) -- (0,1.5) -- (0,0.5)-- (1,0.5);
			
			\filldraw[thick,lightgray] (0,-1.5) -- (1,-1.5) -- (1,-0.5) -- (0,-0.5) -- (0,-1.5);
			\draw[thick] (0,-1.5) -- (1,-1.5) -- (1,-0.5) -- (0,-0.5) -- (0,-1.5)-- (1,-1.5);
			
			\node[black] at (0.58,1) {$\mathcal{I}_6^{w_1}$};
			\node[black] at (0.58,-1) {$\mathcal{I}_6^{w_2}$};
			

		\end{tikzpicture}
		\caption{The seven-point integral $\CI_7$ can be seen as two six-point integrals $\CI_6$, which are connected through the central emission vertex $C$.}
		\label{fig:6pt-7pt}
	\end{center}
\end{figure}

When further comparing the core functionals $\CI_7$ and $\CI_6$ the structural similarities are suggestive for testing, whether it is possible to write $\CI_7$ as a combination of two $\CI_6$ integrals linked by the central emission vertex as shown in \figref{fig:6pt-7pt}. Indeed, this is possible \cite{DelDuca:2019tur} and one finds
\begin{align}
		&\CI_7^{w_1,w_2}[\CF(\nu,n,\mu,m)]=
		\sum_{n=-\infty}^\infty\sum_{m=-\infty}^\infty\int_{-\infty}^\infty \frac{d\nu}{2\pi}\int_{-\infty}^\infty \frac{d\mu}{2\pi}w_1^{i\nu{+}\frac{n}2}(w_1^*)^{i\nu{-}\frac{n}2}w_2^{i\mu{+}\frac{m}2}(w_2^*)^{i\mu{-}\frac{m}2}\nonumber\\
		&\hspace{50ex}\times (-1)^{n+m}\tilde{C}(\nu,n,\mu,m)\,\CF(\nu,n,\mu,m)\notag\\
		&\quad=\sum_{n=-\infty}^\infty (-1)^n\left(\frac{w_1}{w_1^*}\right)^{\frac{n}2}\hspace{-1ex}\int_{-\infty}^\infty \frac{d\nu}{2\pi}|w_1|^{2i\nu}\CI_6^{w_2}\left[\frac{\mu^2+\frac{m^2}4}{2}\tilde{C}(\nu,n,\mu,m)\,\CF(\nu,n,\mu,m)\right]\notag\\
		&\quad=\CI_6^{w_1}\left[\frac{\nu^2+\frac{n^2}4}{2}\,\CI_6^{w_2}\left[\frac{\mu^2+\frac{m^2}4}{2}\tilde{C}(\nu,n,\mu,m)\,\CF(\nu,n,\mu,m)\right]\right].
\end{align}
Note that the definition of $\CI_7$ features no factors of $\nu^2+\frac{n^2}4$ and $\mu^2+\frac{m^2}4$ in the integrand in contrast to the definition of $\CI_6$ in \eqn{eqn:R6shorthand}. So in order to achieve the above connection between $\CI_6$ and $\CI_7$ these factors are inserted manually. The superscripts on the core functionals denote which anharmonic ratio $w_i$ belongs to them. Last, we defined $\tilde{C}$ as the impact factor $C$ but without the factor $(-1)^{n+m}$, such that
\begin{equation}
	C(\nu,n,\mu,m)=(-1)^{n+m}\tilde{C}(\nu,n,\mu,m).
\end{equation}
Thus, the results for the six-point remainder function obtained in \secref{sec:doublecopy} bear the hope of being able to be generalized to the seven-point remainder function, because they share several structures. In particular can $\CI_7$ be considered as an iterated integration of $\CI_6$. 

Explicitly we can do this by using the analogue of the transformation \eqn{eqn:changeofvariables} from section~\ref{sec:doublecopy} for two sets of variables $(w_1,\nu,n)$, $(w_2,\mu,m)$:
\begin{equation}
	\pm\chi_{1\mp}=i\nu\pm\frac{n}2\,,\quad \pm\chi_{2\mp}=i\mu\pm\frac{m}2\,.
\end{equation}
In the new variables, $\CI_7$ can be represented as nine terms (where again analytic continuation of the absolute values has to be considered):
\begin{align}
	\label{eq:7pt2}
		\CI_7^{w_1,w_2}[E_{\nu,n}^kE_{\mu,m}^{\ell-k-1}]
		&=\frac{1}{16\pi^4}\int d^2\chi_{1\pm}(w_1^*)^{-\chi_{1+}}w_1^{\chi_{1-}}E_{\nu,n}^k\Gammafn_1\int d^2\chi_{2\pm}(w_2^*)^{-\chi_{2+}}w_2^{\chi_{2-}}E_{\mu,m}^{\ell-k-1}\Gammafn_2 \tilde{C}\nonumber\\
		&\qquad\qquad+(w_1{\leftrightarrow} w_1^*)+(w_2{\leftrightarrow} w_2^*)+\big((w_1,w_2){\leftrightarrow} (w_1^*,w_2^*)\big)\nonumber\\
		&\quad+\frac{1}{8\pi^3}\int d^2\chi_{1\pm}(w_1^*)^{-\chi_{1+}}w_1^{\chi_{1-}}E_{\nu,n}^k\Gammafn_1\int_{-\infty}^{\infty}d\mu\,|w_2|^{2i\mu}\frac{E_{\mu,0}^{\ell-k-1}}{\mu^2}\tilde{C}+(w_1{\leftrightarrow} w_1^*)\nonumber\\
		&\quad+\frac{1}{8\pi^3}\int_{-\infty}^{\infty}d\nu\,|w_1|^{2i\nu}\frac{E_{\nu,0}^k}{\nu^2}\int d^2\chi_{2\pm}(w_2^*)^{-\chi_{2+}}w_2^{\chi_{2-}}E_{\mu,n}^{\ell-k-1}\Gammafn_2\tilde{C}+(w_2{\leftrightarrow} w_2^*)\nonumber\\
		&\quad+\frac1{4\pi^2}\int_{-\infty}^{\infty}d\nu\,|w_1|^{2i\nu}\frac{E_{\nu,0}^k}{\nu^2}\int_{-\infty}^{\infty}d\mu\,|w_2|^{2i\mu}\frac{E_{\mu,0}^{\ell-k-1}}{\mu^2}\tilde{C}\,.
\end{align}
The integration path in this formula is parallel to the imaginary axis stretching from $1/4-i\infty$ to $1/4+i\infty$ and we defined $\Gammafn_i:=\Gammafn(1{-}\chi_{i+}{-}\chi_{i-})\Gammafn(\chi_{i+}{+}\chi_{i-})$. The central emission vertex $\tilde{C}$ takes in the new variables the form
\begin{equation}
	\tilde{C}=\frac{\Gammafn(-\chi_{1-})\Gammafn(\chi_{2-})\Gammafn(-\chi_{1+}+\chi_{2+})}{\Gammafn(1-\chi_{1+})\Gammafn(1+\chi_{2+})\Gammafn(1-\chi_{1-}+\chi_{2-})}\,.
\end{equation}

There is three distinct types of terms in \eqn{eq:7pt2}: four terms exhibiting four integrations, four terms with three integrations and finally one term with just two integrations.

Employing the two-dimensional inverse Mellin transform $\CM_2^{-1}$ of a two-variable function
\begin{equation}
	\CM_2^{-1}[\CF(\chi_1,\chi_2)](x,y):=\int_{c-i\infty}^{c+i\infty}\frac{d\chi_1}{2\pi i}\int_{c-i\infty}^{c+i\infty}\frac{d\chi_2}{2\pi i}x^{-\chi_1}y^{-\chi_2}\CF(\chi_1,\chi_2),
\end{equation}
one can repeat the steps for $\CI_6$ from section \ref{sec:doublecopy} both in the $\chi_{1\pm}$ and the $\chi_{2\pm}$ variables to write the first term of \eqn{eq:7pt2} as\footnote{We omit the $\pm$-indices in this expression: the first inverse Mellin transform has the $\chi_+$ and the second one the $\chi_-$.}
\begin{align}
		&\int d^2\chi_{1\pm}(w_1^*)^{-\chi_{1+}}w_1^{\chi_{1-}}E_{\nu,n}^k\Gammafn_1
		\int d^2\chi_{2\pm}(w_2^*)^{-\chi_{2+}}w_2^{\chi_{2-}}E_{\mu,m}^{\ell-k-1}\Gammafn_2 \tilde{C}\nonumber\\
		&\quad\quad\quad\propto \sum_{i,j}\int_{0}^{w_1}\frac{du}{u+1}\int_{0}^{w_2}\frac{dv}{v+1}\,\CM_2^{-1}\left[f_i(\chi_1)\tilde{f}_j(\chi_2)\frac{\Gammafn(-\chi_1+\chi_2)}{\Gammafn(1-\chi_1)\Gammafn(1+\chi_2)}\right](u w_1^*,v w_2^*)\nonumber\\
		&\quad\quad\hspace{30ex}\times\CM_2^{-1}\left[g_i(\chi_1)\tilde{g}_j(\chi_2)\frac{\Gammafn(-\chi_1)\Gammafn(\chi_2)}{\Gammafn(1-\chi_1+\chi_2)}\right](u/w_1,v/w_2).
\end{align}
In the above integrals the contour chosen is parallel to imaginary axis at a distance of $c=1/4$. In analogy to \eqn{eq:Fseparated} for the six-point case, we defined
\begin{equation}
	E_{\nu,n}^k(\xp,\xm)=:\sum_i f_i(\xp)g_i(\xm),\quad\text{and}\quad E_{\mu,m}^{\ell-k-1}(\xp,\xm)=:\sum_j \tilde{f}_i(\xp)\tilde{g}_i(\xm).
\end{equation}
Similar manipulations can be applied to the three other terms in the first block of \eqn{eq:7pt2}. 

The next four terms (with three integrations in each term) can be re-written as
\begin{align}
		&\int d^2\chi_{1\pm}(w_1^*)^{-\chi_{1+}}w_1^{\chi_{1-}}E_{\nu,n}^k\Gammafn_1\int_{-\infty}^{\infty}d\mu\,|w_2|^{2i\mu}\frac{E_{\mu,0}^{\ell-k-1}}{\mu^2}\tilde{C}\notag\\
		&\quad \propto \sum_i\int_{0}^{w_1}\frac{du}{u+1}\CM^{-1}\Bigg[\frac{E_{i\mu,0}^{\ell-k-1}}{\mu^2}\frac{\Gammafn(-\mu)}{\Gammafn(1+\mu)}\CM^{-1}\left[f_i(\xp)\frac{\Gammafn(-\xp+\mu)}{\Gammafn(1-\xp)}\right](u w_1^*)\notag\\
		&\hspace{34ex}\times\,\CM^{-1}\left[g_i(\xm)\frac{\Gammafn(-\xm)}{\Gammafn(1-\xm-\mu)}\right](u/ w_1)\Bigg]\left(|w_2|^2\right),
\end{align}
where we substituted $i\mu\rightarrow-\mu$ for the outermost inverse Mellin transform in $\mu$: this leads to a vanishing offset of the integration contour from the imaginary axis is $c=0$. Note, that this integration runs over a pole at $\mu=0$, so that this integration has to be regarded as a principal value, i.e.~the $\mu$-integration is calculated over the imaginary axis, excluding an $\ve$-neighborhood around the pole and taking the limit $\varepsilon\rightarrow0$ after the integration.

Finally, the last term in \eqn{eq:7pt2} can be recast into the form
\begin{align}
		&\int_{-\infty}^{\infty}d\nu\,|w_1|^{2i\nu}\frac{E_{\nu,0}^k}{\nu^2}\int_{-\infty}^{\infty}d\mu\,|w_2|^{2i\mu}\frac{E_{\mu,0}^{\ell-k-1}}{\mu^2}\tilde{C}\notag\\
		&\quad\propto \CM_2^{-1}\left[\frac{E_{i\nu,0}^k}{\nu^2}\frac{E_{i\mu,0}^{\ell-k-1}}{\mu^2}\frac{\Gammafn(\nu)\Gammafn(-\mu)\Gammafn(\mu-\nu)}{\Gammafn(1-\nu)\Gammafn(1+\mu)\Gammafn(1+\nu-\mu)}\right]\left(|w_1|^2,|w_2|^2\right),
\end{align}
where we again substituted $i\nu\rightarrow-\nu$, $i\mu\rightarrow-\mu$ and $c=0$ for both inverse Mellin transforms. Again, the integrals are to be interpreted as principal values.
From the current form of the seven-point remainder function it is not clear to us how to proceed in order to disentangle the four integrations. However, it would be disappointing if this problem could not be solved: probably the set of transformations is just too intricate for the current time.   


\section{Conclusion and outlook}
\label{sec:results}
In the current article we reported on the decomposition of a class of single-valued observables -- the remainder function in \Nfour sYM theory -- into a KLT-square of multi-valued objects, whose structure resembles open-string or gauge theory amplitudes. While not yet associated to a particular theory, the objects on the open/gauge-theory side belong to the class of Mellin transforms of polygamma functions.

Our approach works for the simplest scenario, the six-point remainder function. For the seven-point function we have discussed the setup and a couple of steps, but failed to disentangle the iterated integrations completely. The main problem is the simultaneous disentangling of various iterated integrals in several sets of variables. 

In our calculations, we start from the single-valued object in order to obtain products of multi-valued objects. This algorithmic approach is shared by the original KLT construction \cite{KLT} and Stieberger`s generalization to genus one \cite{Stieberger:2022lss}. 

Double-copy (or single-valued) constructions can be divided into symmetric and asymmetric versions: while the single-valued construction of Brown and Dupont \cite{BrownDupont1,BrownDupont2} is clearly asymmetric by giving preference in simplicity to the holomorphic side of the multi-valued objects, the KLT relation is symmetric in both double-copy components. Both approaches naturally have their virtues, however, decomposing a closed-side object will likely lead to symmetric versions.  

Since we hope that every single-valued object with total symmetry in the arguments can be decomposed in a meaningful way into multi-valued objects with ordered arguments, let us point out a couple of questions to be answered in the future:  
\begin{itemize}
	\item It would be nice to perform the disentangling for the seven-point amplitude, which should pave a combinatorial way for a general formalism. 
	\item Once the step from six to seven points is understood, one could hope to recover the recursive patterns based on matrix manipulations put forward in \rcite{DelDuca:2019tur}. However, currently the languages are rather far apart and therefore several translations might be necessary. 
	\item The largest question remaining unanswered is, whether all theories producing observables with only single-valued objects or coefficients can be decomposed such that the reverse process yields a double-copy. In order to substantiate our hope, it would be favorable to have more decompositions of theories with single-valued observables on our disposal.
\end{itemize}
%


\subsection*{Acknowledgments}

We are grateful to Claude Duhr and Stephan Stieberger for discussions about the content of this article and related subjects and to Egor Im and Beat Nairz for comments on the draft. The work of JB and KB is partially supported by the Swiss National Science Foundation through the NCCR SwissMAP.


\section*{Appendix}
\appendix

\section{Examples of double-copy representation for various integrals}
\label{app:calculations}
In section \ref{sec:separating} it was shown that the formula \eqn{eq:R6full} for the remainder function $R_6$ can be rewritten as a double-copy with an outer scaling integral. In terms of the core functional $\CI_6[\CF]$ introduced in section \ref{sec:doublecopy}, validity of the formula
\begin{align}
	\label{eq:shorthand_final}
	\CI_6[\CF]=&-2\sum_i\int_0^w\frac{du}{u+1}\CM^{-1}\left[\frac{f_i(\xp)}{\xp}\right](uw^*)\CM^{-1}\left[\frac{g_i(\xm)}{\xm}\right]\left(\frac{u}{w}\right)\notag\\
		&+(w\leftrightarrow w^*)+\frac{1}{\pi}\int_{-\infty}^\infty d\nu |w|^{2i\nu}\frac{\CF(\nu,0)}{\nu^2}
\end{align}
was proven. In this appendix the corresponding calculations for $\CI_6[1]$, $\CI_6[V]$, $\CI_6[N]$ and $\CI_6[E_\psi]$ will be presented, discussed and compared with available results in the literature. Several results of the core functional $\CI_6[\CF]$ for various arguments $\CF$ at higher weights have been checked numerically and reproduce available results in the literature.

In preparation for the calculation, let us collect a couple of identities for inverse Mellin transformations. As described in \secref{sec:separating}, depending on the value of the argument transform, the integration contour is either to be closed on the left or on the right side of the complex plane, resulting in contributions from poles on the left hand side of the integration contour exclusively. Using this prescription, the two identities 
\begin{subequations}
\begin{align}
	\label{eq:res1}
	\CM^{-1}\left[\frac{1}{\chi^k}\right](u)&=\left\{\begin{array}{ll}\frac{(-1)^{k-1}}{(k-1)!}\log^{k-1}(u), & 0<u<1\\ 0, & u>1\end{array}\right.\\
	\label{eq:res2}
	\CM^{-1}\left[\frac1{\chi}(\psi(1+\chi)+\gamma_E)\right](u)&=\left\{\begin{array}{ll}-\log(1-u), & 0<u<1\\ 0, & u>1\end{array}\right.
\end{align}
\end{subequations}
are valid, where we have used the condition $\operatorname{Re}\chi_\pm=1/4>0$ in our case. With these two identities, we can now proceed with the calculations of the core functional.

\subsection{\texorpdfstring{$\CI_6[1]$}{I6[1]}}
For $\CF=1$, we find $f_1=g_1=1$ and use \eqn{eq:res1} for the case $k=1$. The ($n=0$)-term (i.e.~the last term of the identity \eqn{eq:shorthand_final}) gives $\CG_0(w)=\log|w|^2$ (this comes from half the pole at $n=\nu=0$, see discussions in refs.~\cite{BroedelSprenger,Dixon:2012yy}). We arrive at
\begin{align}
		\CI_6[1]&=-2\int_0^w\frac{du}{u+1}\underbrace{\CM^{-1}\left[\frac{1}{\xp}\right](uw^*)}_{=1}\underbrace{\CM^{-1}\left[\frac{1}{\xm}\right]\left(\frac{u}{w}\right)}_{=1}+(w\leftrightarrow w^*)+\underbrace{\frac{1}{\pi}\int_{-\infty}^\infty d\nu |w|^{2i\nu}\frac{\CF(\nu,0)}{\nu^2}}_{=\CG_0}\notag\\
		&=-2\underbrace{\underbrace{\int_0^w\frac{du}{u+1}}_{=\log(w+1)}+\underbrace{(w\leftrightarrow w^*)}_{=\log(w^*+1)}}_{=\log|w+1|^2=\CG_1(w)}+\CG_0\notag\\
		&=-2\CG_1(w)+\CG_0(w)\,,
\end{align}
which agrees with the known results for $\CI_6[1]$ \cite{BroedelSprenger,Dixon:2012yy}.
	
\subsection{\texorpdfstring{$\CI_6[V]$}{I6[V]}}
	For $\CF=V=\frac{1}{2\xp}-\frac1{2\xm}$ we have $f_1=\frac1{2\xp}$, $g_1=1$, $f_2=1$, $g_2=-\frac1{2\xm}$. The ($n=0$)-term gives $-\CG_{00}(w)=-\frac12\log^2|w|^2=-\frac12\CG_0(w)^2$ (again coming from half of the residue at $n=\nu=0$). Now we use \eqn{eq:res1} for $k=2$ and obtain
\begin{align}
	\CI_6[V]&=-2\int_0^w\frac{du}{u+1}\underbrace{\CM^{-1}\left[\frac{1}{2\chi^2_+}\right](uw^*)}_{=-\frac12\log(uw^*)}\underbrace{\CM^{-1}\left[\frac{1}{\xm}\right]\left(\frac{u}{w}\right)}_{=1}\notag\\
	&\quad-2\int_0^w\frac{du}{u+1}\underbrace{\CM^{-1}\left[\frac{1}{\xp}\right](uw^*)}_{=1}\underbrace{\CM^{-1}\left[-\frac{1}{2\chi^2_-}\right]\left(\frac{u}{w}\right)}_{=\frac12\log(u/w)}+(w\leftrightarrow w^*)+\underbrace{\frac{1}{\pi}\int_{-\infty}^\infty d\nu |w|^{2i\nu}\frac{V}{\nu^2}}_{=-\frac12\CG_0(w)^2}\notag\\
	&=\int_0^w\frac{du}{u+1}\underbrace{\left(\log(uw^*)-\log(u/w)\right)}_{=\log(ww^*)=\CG_0(w)}+(w\leftrightarrow w^*)-\frac12\CG_0(w)^2\notag\\
	&=\CG_0(w)\underbrace{\int_0^w\frac{du}{u+1}}_{=\log(w+1)}+(w\leftrightarrow w^*)-\frac12\CG_0(w)^2\notag\\
	&=\CG_0(w)\underbrace{(\log(w+1)+(w\leftrightarrow w^*))}_{=\log|w+1|^2=\CG_1(w)}-\frac12\CG_0(w)^2\notag\\
	&=\CG_0(w)\left(\CG_1(w)-\frac12\CG_0(w)\right).
\end{align}
This matches the result for $\CI_6[V]$ in the literature \cite{Dixon:2012yy}.
	
\subsection{\texorpdfstring{$\CI_6[N]$}{I6[N]}}
For $\CF=N=\frac{1}{\xp}+\frac1{\xm}$ we have $f_1=\frac1{\xp}$, $g_1=1$, $f_2=1$, $g_2=\frac1{\xm}$. Furthermore, the ($n=0$)-term vanishes, because $N|_{n=0}=0$. For $\CF=N$ we need to slightly adjust the formula \eqn{eq:shorthand_final}, as due to its symmetry, $N$ does not belong to the functions appearing in the expansion of the remainder function (see section \ref{sec:6pt} for a discussion). Because the $(w\leftrightarrow w^*)$-term comes from the negative $n$ in the original formulation of the remainder function, it should come with a minus sign, because our formalism (see \secref{sec:doublecopy}) maps $n$ to $|n|$. Here, we again use the case $k=2$ of \eqn{eq:res1}, and obtain\footnote{We can exchange $uw^*,u/w\in\mathbb{R}_+$ for $|uw^*|,|u/w|$ in the third line of the equation.}
\begin{align}
			\CI_6[N]&=-2\int_0^w\frac{du}{u+1}\underbrace{\CM^{-1}\left[\frac{1}{\chi^2_+}\right](uw^*)}_{=-\log(uw^*)}\underbrace{\CM^{-1}\left[\frac{1}{\xm}\right]\left(\frac{u}{w}\right)}_{=1}\notag\\
			&\quad-2\int_0^w\frac{du}{u+1}\underbrace{\CM^{-1}\left[\frac{1}{\xp}\right](uw^*)}_{=1}\underbrace{\CM^{-1}\left[\frac{1}{\chi^2_-}\right]\left(\frac{u}{w}\right)}_{=-\log(u/w)}-(w\leftrightarrow w^*)+\underbrace{\frac{1}{\pi}\int_{-\infty}^\infty d\nu |w|^{2i\nu}\frac{N}{\nu^2}}_{=0}\notag\\
			&=2\int_0^w\frac{du}{u+1}\underbrace{\left(\underbrace{\log(uw^*)}_{=\log|uw^*|}+\underbrace{\log(u/w)}_{=\log|u/w|}\right)}_{=\log|u|^2=\CG_0(u)}-(w\leftrightarrow w^*)\notag\\
			&=2\int_0^w\frac{du}{u+1}\CG_0(u)-2\int_0^{w^*}\frac{du}{u+1}\CG_0(u)\notag\\
			&=2\CG_{10}(w)-2\CG_{01}(w).
\end{align}
In the last line we used the defining differential equation for svMPLs (see \eqn{eq:svhpl_gen_func}). The result for $\CI_6[N]$ again agrees with the literature \cite{Dixon:2012yy}. Equivalently one could write out the third line in terms of MPLs and recognize them as svMPLs.
	
\subsection{\texorpdfstring{$\CI_6[E_{\nu,n}]$}{I6[E(nu,n)}}
Expressions containing polygamma factors need special attention in this formalism. We express $\CF=E_{\nu,n}=-|N|/2+E_\psi=-\frac1{2\xp}-\frac1{2\xm}+\psi(1+\xp)+\gamma_E+\psi(1+\xm)+\gamma_E$ through $f_1=-\frac1{2\xp}$, $g_1=1$, $f_2=1$, $g_2=-\frac1{2\xm}$, $f_3=\psi(1+\xp)+\gamma_E$, $g_3=1$, $f_4=1$, $g_4=\psi(1+\xm)+\gamma_E$. The ($n=0$)-term gives the MPL $-2G_{01}(|w|^2)$ (here the residue at $n=\nu=0$ vanishes but we have an infinite tower of residues from the digamma function at $\nu=-n$, $n\in\mathbb{N}$). We now use \eqn{eq:res2} and get\footnote{Unlike for $\CI_6[N]$ the sign of the $(w\leftrightarrow w^*)$-term is unaltered, since we have $|N|$ here.}
\begin{align}
		\label{eq:e_psi}
			\CI_6[E_{\nu,n}]&=-2\int_0^w\frac{du}{u+1}\underbrace{\CM^{-1}\left[\frac{1}{\chi^2_+}\right](uw^*)}_{=-\log(uw^*)}\underbrace{\CM^{-1}\left[\frac{1}{\xm}\right]\left(\frac{u}{w}\right)}_{=1}\notag\\
			&\quad-2\int_0^w\frac{du}{u+1}\underbrace{\CM^{-1}\left[\frac{1}{\xp}\right](uw^*)}_{=1}\underbrace{\CM^{-1}\left[\frac{1}{\chi^2_-}\right]\left(\frac{u}{w}\right)}_{=-\log(u/w)}\notag\\
			&\quad-2\int_0^w\frac{du}{u+1}\underbrace{\CM^{-1}\left[\frac{\psi(1+\xp)+\gamma_E}{\xp}\right](uw^*)}_{=-\log(1-uw^*)}\underbrace{\CM^{-1}\left[\frac{1}{\xm}\right]\left(\frac{u}{w}\right)}_{=1}\notag\\
			&\quad-2\int_0^w\frac{du}{u+1}\underbrace{\CM^{-1}\left[\frac{1}{\xp}\right](uw^*)}_{=1}\underbrace{\CM^{-1}\left[\frac{\psi(1+\xm)+\gamma_E}{\xm}\right]\left(\frac{u}{w}\right)}_{=-\log(1-u/w)}\notag\\
			&\quad+(w\leftrightarrow w^*)+\underbrace{\frac{1}{\pi}\int_{-\infty}^\infty d\nu |w|^{2i\nu}\frac{E_{\nu,n}}{\nu^2}}_{=-2G_{01}(|w|^2)}\notag\\
			&=-G_{10}-\overline{G}_{10}+G_{01}+\overline{G}_{01}-G_0\overline{G}_1-\overline{G}_0G_1\notag\\
			&\quad+2\int_0^w\frac{du}{u+1}\left(\log(1-uw^*)+\log(1-u/w)\right)+(w\leftrightarrow w^*)-2G_{01}(|w|^2).
\end{align}
Here, the MPLs are evaluated at $-w$ if not stated otherwise (and at $-w^*$ for $\overline{G}$). To show that this result is the same as the svMPL expression $\CI_6[E_{\nu,n}]=2\CG_{11}-\CG_{01}-\CG_{10}$ known from \rcite{Dixon:2012yy}, we need the two identities 
\begin{subequations}
\begin{align}
	\label{eq:HIdentity1}
	G_{01}(z)-G_{11}(z)&=-\int_0^{z}\frac{du}{u-1}\log\left(1-u/z\right),\\
	\label{eq:HIdentity2}
	G_1(z)G_1(\bar{z})&=\int_0^z\frac{du}{u-1}\log(1-u\bar{z})+(z\leftrightarrow \bar{z})-G_{01}(|z|^2),
\end{align}
\end{subequations}
and the identification $(-w,-w^*)=(z,\bar{z})$:
\begin{equation}
	\CI_6[E_{\nu,n}]=2\CG_{11}-\CG_{01}-\CG_{10}=-G_{01}-G_{10}-G_0\overline{G}_1-\overline{G}_0G_1-\overline{G}_{01}-\overline{G}_{10}+2G_{11}+2\overline{G}_{11}+2G_1\overline{G}_1.
\end{equation}
The first identity \eqn{eq:HIdentity1} for Goncharov polylogarithms $G$ \cite{Goncharov:2001iea} can be shown using the identity (5.21) from ref.~\cite{Broedel:2013},
\begin{align}
	\int_0^{z}\frac{du}{u-1}\log\left(1-u/z\right)&\stackrel{u\rightarrow\frac{u}{z}}{=}\int_0^1\frac{du}{u-1/z}\log(1-u)\notag\\
	&=G(1/z,1;1)=G(1,z;z)\notag\\
	&\stackrel{(5.21)\text{ \cite{Broedel:2013}}}{=}G_{11}(z)-G_{01}(z)\,.
\end{align}
	The second identity, \eqn{eq:HIdentity2}, has been checked numerically on the unit disk.


\section{Convergence proofs}
\label{app:proof}
This appendix provides proofs for the vanishing of contributions from certain stretches of integration contours in particular limits. These arguments have been used in \secref{sec:doublecopy}.

\subsection{Closing the Sommerfeld--Watson contour }
\label{app:proof1}
In section \ref{sec:doubleint} the integration contour for $z$ is closed in the complex upper half-plane as the radius of the contour goes to infinity. Here we will give the proof for relation \eqref{eq:412firsteqn} (repeated for convenience):
\begin{equation}
	\lim_{R\rightarrow\infty}\left(\int_{-R}^R f(z)dz+\int_{\gamma_R}\!\!\!f(z)dz\right)=2\pi i\sum_{n\in\mathbb{N}}\operatorname{Res}_{i\left(\frac{2n-1}{2}\right)}f=\sum_{n\in\mathbb{Z}\backslash\{0\}}\!\!(-1)^n\left(\frac{w}{w^*}\right)^{\frac{n}2}\frac{\CF(\nu,n)}{\nu^2+\frac{n^2}2}\,,
\end{equation}
where $\gamma_R(t)=Re^{it},\,t\in[0,\pi],\,R\in\mathbb{N}$.

\paragraph{Proof.} Let $1>\varepsilon>0$ and consider $\nu\in\mathbb{R}$ fixed. We will show that by choosing the radius $R$ of the integral contour very large, the deviation of the integral from the infinite sum from \eqn{eq:goal} (up to the ($n=0$)-term discussed in \secref{sec:doubleint}) is smaller than $\varepsilon$:
\begin{equation}
	\left|\sum_{n\in\mathbb{Z}\backslash\{0\}}(-1)^n\left(\frac{w}{w^*}\right)^{\frac{n}2}\frac{1}{\nu^2+\frac{n^2}4}-\int_{-R}^Rf(z)\,dz\right|<\varepsilon\,.
\end{equation}
First, we note that when choosing a radius $R=n_0\in\mathbb{N}$ for the contour we can estimate the terms of the sum from poles that are not contained in the contour from above by
\begin{align}
	\left|\sum_{n=n_0}^\infty(-1)^n\left(\left(\frac{w}{w^*}\right)^{\frac{n}2}+\left(\frac{w}{w^*}\right)^{-\frac{n}2}\right)\frac{1}{\nu^2+\frac{n^2}4}\right|&\leq \sum_{n=n_0}^\infty\underbrace{\left|\left(\frac{w}{w^*}\right)^{\frac{n}2}+\left(\frac{w}{w^*}\right)^{-\frac{n}2}\right|}_{\leq 2}\left|\frac{1}{\nu^2+\frac{n^2}4}\right|\notag\\
	&\leq2\sum_{n=n_0}^\infty\underbrace{\frac{1}{\nu^2+\frac{n^2}4}}_{\leq\frac{4}{n^2}}\leq8\sum_{n=n_0}^\infty\frac{1}{n^2}\,,
\end{align}
where we used that $\frac{w}{w^*}$ is just a complex phase (with absolute value $1$). The last term denotes the remaining parts of the Basel problem \cite{Basel2}, of which we know that it converges (against $\pi^2/6$), meaning that when choosing $n_0$ appropriately large the sum of the terms for $n\geq n_0$ can be made arbitrarily small. Thus, there exists an $n_0\in\mathbb{N}$ such that
\begin{equation}
\label{eq:first_est}
	\left|\sum_{n=n_0}^\infty(-1)^n\left(\left(\frac{w}{w^*}\right)^{\frac{n}2}+\left(\frac{w}{w^*}\right)^{-\frac{n}2}\right)\frac{1}{\nu^2+\frac{n^2}4}\right|<\frac{\varepsilon}2\,.
\end{equation}
Next, we want to estimate the absolute value of the contour of the semicircle in the upper complex plane. For this let $R=n_1\in\mathbb{N}$ with\footnote{Since $\nu$ is fixed and the radius can always be chosen sufficiently large, we can use $n_1>4\nu$, so that by increasing $n_1$ the absolute value of the denominator of $X_2$ and of terms in $\CF$ will certainly only increase.} $n_1>4\nu$. With $\gamma_{n_1}$ the path as in \eqn{eq:contour} we can estimate
\begin{align}
	\left|\int_{\gamma_{n_1}}f(z)\,dz\right|&\leq\underbrace{\pi n_1}_{\text{length of }\gamma_{n_1}}\max_{t\in[0,\pi]}\left|f(\gamma_{n_1}(t))\right|\notag\\
	&=\pi n_1\max_{t\in[0,\pi]}\left|\frac{\cosh\left(\left(n_1e^{it}+\frac{i}2\right)\phi\right)}{\cosh(\pi n_1e^{it})}\right|\left|\frac{\CF\left(\nu,-in_1e^{it}+\frac12\right)}{\nu^2-\frac{\left(n_1e^{it}+\frac{i}2\right)^2}{4}}\right|\notag\\
	&\leq\pi n_1\max_{t\in[0,\pi]}\underbrace{\left|\frac{\cosh\left(\left(n_1e^{it}+\frac{i}2\right)\phi\right)}{\cosh(\pi n_1e^{it})}\right|}_{=:X_1(t,n_1)}\max_{t'\in[0,\pi]}\underbrace{\left|\frac{\CF\left(\nu,-in_1e^{it'}+\frac12\right)}{\nu^2-\frac{\left(n_1e^{it'}+\frac{i}2\right)^2}{4}}\right|}_{=:X_2(t',n_1)}.
\end{align}
In order to find an estimate for $X_1$, we distinguish two cases:
\begin{itemize}
	\item for $t=\frac{\pi}{2}$, we calculate
	\begin{align}
		X_1\left(t\,{=}\frac{\pi}2,n_1\right)=\left|\frac{\cosh\left[\left(n_1+\frac{1}2\right)i\phi\right]}{\cosh(i\pi n_1)}\right|
		=\left|\frac{\cos\left[\left(n_1+\frac{1}2\right)\phi\right]}{\cos(\pi n_1)}\right|
		\leq 1.
	\end{align}
	\item for $t\neq\frac{\pi}2$, we consider the behavior of $\cosh(z)$ for $z=x+iy\in\mathbb{C}$ and $|x|$ large\footnote{Note that for $t\neq\frac{\pi}2$ the arguments of our $\cosh$ always have a real part that becomes large if $n_1$ becomes large.}:
	\begin{equation}
		|\cosh(x+iy)|=\frac12\left|e^xe^{iy}+e^{-x}e^{-iy}\right|\sim\frac12\left|e^{|x|}e^{i\operatorname{sign}(x)y}\right|=\frac12e^{|x|}\,.
	\end{equation}
	Using the above approximation, we can identify the behavior for large $n_1$. One finds
	\begin{equation}
		X_1\left(t\neq\frac{\pi}{2},n_1\right)\sim\frac{e^{n_1|\phi\,\cos t|}}{e^{n_1\pi |\cos t|}}\,,
	\end{equation}
	such that for $n_1\rightarrow\infty$
	\begin{equation}
		\begin{cases}
			X_1\rightarrow0 & \text{if } |\phi|\neq\pi\,,\\
			X_1\rightarrow1 & \text{if } |\phi|=\pi\,.
		\end{cases}
	\end{equation}
\end{itemize}
In conclusion, there exists an $n_1'\in\mathbb{N}$ such that $X_1(t,n_1)\leq1+\epsilon$ for all $n_1>n_1'$.

The part $X_2$ converges to zero for large $n_1$ at least asymptotically\footnote{For the possible terms in $\CF$ see the previous subsection. First, check that they indeed have at least an asymptotic $\sim \frac{1}{n_1}$ convergence to zero behavior for large $n_1$ if they contain at least one $V$ or $N$. The polygamma functions have asymptotics $\psi^{(k)}(z)=\mathcal{O}(1/z)$ for $|z|\rightarrow\infty$ \cite{NIST}. Euler--Mascheroni constants $\gamma_E$ will contribute a constant to the asymptotic. Last, the digamma function $\psi$ itself has the asymptotic $\psi(z)=\log(z)+\mathcal{O}(1/z)$ for $|z|\rightarrow\infty$ \cite{NIST}. Thus, in the worst case we have $\CF=E_\psi^k$ for $k\in\mathbb{N}$, so that in this case $X_2\sim \frac{\log^k(n_1)}{n_1^2}$.} as $X_2\sim \frac{\log^k(n_1)}{n_1^2}$, $k\in\mathbb{N}$ fixed, independent of the value for $t_1$. Thus, there exists an $n_1''$, such that $\pi n_1''X_2(t',n_1'')<\frac{\varepsilon}4$ for any $t'\in[0,\pi]$. Thus, by choosing $n_1=\max\{n_1',n_1''\}$ as described above we can estimate
\begin{equation}
	\left|\int_{\gamma_{n_1}}f(z)\,dz\right|\leq(1+\varepsilon)\frac{\varepsilon}4=\frac{\varepsilon}4+\frac{\varepsilon^2}4<\frac{\varepsilon}2\,.
\end{equation}
Now, by choosing $R=\max\{n_0,n_1\}\in\mathbb{N}$ we can combine this result with \eqn{eq:first_est} to get
\begin{align}
	&\left|\sum_{n\in\mathbb{Z}\backslash\{0\}}(-1)^n\left(\frac{w}{w^*}\right)^{\frac{n}2}\frac{1}{\nu^2+\frac{n^2}4}-\int_{-R}^Rf(z)\,dz\right|\notag\\
	&\hspace{10ex}=\left|\sum_{n=1}^\infty(-1)^n\left(\left(\frac{w}{w^*}\right)^{\frac{n}2}+\left(\frac{w}{w^*}\right)^{-\frac{n}2}\right)\frac{1}{\nu^2+\frac{n^2}4}\right.\notag\\
	&\hspace{25ex}\left.-\sum_{n=1}^{R}(-1)^n\left(\left(\frac{w}{w^*}\right)^{\frac{n}2}+\left(\frac{w}{w^*}\right)^{-\frac{n}2}\right)\frac{1}{\nu^2+\frac{n^2}4}+\int_{\gamma_R}f(z)\,dz\right|\notag\\
	&\hspace{10ex}\leq\left|\sum_{n=R}^\infty(-1)^n\left(\left(\frac{w}{w^*}\right)^{\frac{n}2}+\left(\frac{w}{w^*}\right)^{-\frac{n}2}\right)\frac{1}{\nu^2+\frac{n^2}4}\right|+\left|\int_{\gamma_R}f(z)\,dz\right|<\frac{\varepsilon}2+\frac{\varepsilon}2=\varepsilon\, ,
\end{align}
which concludes the proof of convergence.\qed

\subsection{Contour rotation for the Gamma function identity}
\label{app:proof2}
In \secref{sec:separating} we want to rotate the integration path of the Gamma function identity \eqn{eq:gamma_id} in the complex plane. To prove validity, we need to show (the denoted integration paths can be found in \figref{fig:gamma_id})
\begin{align}
	0=&\hspace{-1ex}\lim_{\begin{array}{c}\scriptstyle R\rightarrow\infty\vspace{-6pt}\\\scriptstyle\varepsilon\rightarrow0\end{array}}\hspace{-1ex}\left[\left(\int_{\gamma_1}{+}\int_{\gamma_R}{+}\int_{\gamma_2}{+}\int_{\gamma_\varepsilon}\right)\frac{du}{u(u+1)}u^{\xp+\xm}\right]\notag\\
	=&\hspace{-1ex}\lim_{\begin{array}{c}\scriptstyle R\rightarrow\infty\vspace{-6pt}\\\scriptstyle\varepsilon\rightarrow0\end{array}}\hspace{-1ex}\left[\int_{\gamma_1}\frac{du}{u(u+1)}u^{\xp+\xm}+\int_{\gamma_2}\frac{du}{u(u+1)}u^{\xp+\xm}\right]\,,
\end{align}
i.e.~we will prove that the contributions from $\gamma_R$ and $\gamma_\varepsilon$ vanish in this limit. Accordingly, the integration path $\gamma_1$ can be replaced by $-\gamma_2$. 

\paragraph{The path $\gamma_R$.}
The contribution from the $\gamma_R$ term can be shown to vanish in the limit $R\rightarrow\infty$ as follows: we fix $\chi_\pm=1/4+ix_\pm$, define $x:=x_++x_-$ and parameterize $\gamma_R(t)=Re^{-it}$, $t\in[0,\phi]$. Assuming $\phi>0$ (the argument works as well for $\phi<0$), one finds
\begin{align}
	\left|\int_{\gamma_R} \frac{du}{u(u+1)}u^{\xp+\xm}\right|&\leq\int_0^\phi\frac{R\, dt}{R|Re^{-it}+1|}R^{1/2}\underbrace{|e^{-it/2}|}_{=1}\underbrace{|e^{ix\log R}|}_{=1}e^{tx}\nonumber\\
	\label{eq:proof}
	&\leq\int_0^\phi\frac{\sqrt{R}\, dt}{\sqrt{R}\sqrt{R-2}}e^{tx}\leq\frac{|\phi|}{\sqrt{R-2}}e^{|\phi x|}\stackrel{R\rightarrow\infty}{\longrightarrow}0\,,
\end{align}
where $|Re^{-it}+1|^2\geq R(R-2)$ was used. 

\paragraph{The path $\gamma_\varepsilon$.}
In a similar way we can show that the $\gamma_\varepsilon$ contribution vanishes in the limit $\varepsilon\rightarrow0$. As in the paragraph above we fix $\chi_\pm=1/4+ix_\pm$ and we parameterize the path $\gamma_\varepsilon(t)=\varepsilon e^{it}$, $t\in[-\phi,0]$ for $0<\varepsilon\ll 1$. Now we estimate
\begin{align}
	\left|\int_{\gamma_\varepsilon}\frac{du}{u(u+1)}u^{\xp+\xm}\right|=&\left|\int_{-\phi}^0\frac{dt\,i\varepsilon e^{it}}{\varepsilon e^{it}(\varepsilon e^{it}+1)}(\varepsilon e^{it})^{\xp+\xm}\right|\notag\\
	\leq& \,\sqrt{\varepsilon}\,|\phi|\max_{t\in[-\phi,0]}\frac{\left|\varepsilon^{i(x_++x_-)}\right|\,\left|e^{it(\xp+\xm)}\right|}{|\varepsilon e^{it}+1|}\notag\\
	\leq& \,\sqrt{\varepsilon}\,|\phi|\max_{t\in[-\phi,0]}\frac{\left|e^{-t(x_++x_-)}\right|}{1-\varepsilon}\stackrel{\varepsilon\rightarrow0}{\longrightarrow}0\,,
\end{align}
where the limit holds for fixed $x_\pm$.

The vanishing of the contributions from $\gamma_\varepsilon$ and $\gamma_R$ implies that the integration path along the real axis in \eqn{eq:gamma_id} can indeed be rotated in the complex plane as depicted in \figref{fig:gamma_id}. \qed


\section{Construction of svHPLs by Brown}
\label{app:conSVHPL}
In this appendix we review the construction \textit{single-valued harmonic polylogarithms} (svHPLs) following \rcite{BrownSVHPL}. As pointed out in \secref{sec:polylogszeta}, they differ from the svMPLs defined in \eqn{eqn:svmpl} merely by minus signs (cf.~\eqn{eqn:svmplrelation}). As a starting point, let us paraphrase the central theorem from \rcite{BrownSVHPL}:
\begin{theorem*}[Brown \cite{BrownSVHPL}]
Let $S=\mathbb{P}^1(\mathbb{C})\backslash\{0,1,\infty\}$ be the punctured complex projective line, and $\mathcal{O}=\mathcal{C}[z,\frac1z,\frac1{1-z}]$ the ring of regular functions on $S$. Then there exists a unique family of single-valued functions 
\begin{equation}
{\{\CL_w(z):w\in X^\times,z\in S\}}\,, 
\end{equation}
each of which is an explicit linear combination of the functions $\operatorname{Li}_w(z)\operatorname{Li}_{w'}(\overline{z})$ where $w,w'\in X^\times$, which satisfy the differential equations:
\begin{equation}
			\frac{\partial}{\partial z}\CL_{x_0w}(z)=\frac{\CL_w(z)}{z},\quad \frac{\partial}{\partial z}\CL_{x_1w}(z)=\frac{\CL_w(z)}{1-z},
\end{equation}
such that $\CL_e(z)=1,\,\CL_{x_0^n}(z)=\frac{1}{n!}\log^n|z|^2$ for all $n\in\mathbb{N}$ and $\lim_{z\rightarrow0}\CL_w(z)=0$ if $w$ is not of the form $x_0^n$. The functions $\CL_w(z)$ satisfy the shuffle relations, and are linearly independent over $\mathcal{O}\overline{\mathcal{O}}$. Every linear combination of the functions $\operatorname{Li}_w(z)\operatorname{Li}_{w'} (\overline{z})$, where $w, w'\in X^\times$, which is single-valued, can be written as a unique linear combination of functions $\CL_w(z)$.
\end{theorem*}
The shuffle relation for svHPLs reads
\begin{equation}
	\CL_{w_1}(z)\CL_{w_2}(z)=\sum_{w\in w_1\shuffle w_2}\CL_{w}(z).
\end{equation}
For the explicit construction of svHPLs, we use so-called generating series, which we express in terms of harmonic polylogarithms (HPLs) \cite{Remiddi:2000}. This explicit construction is discussed for example in ref.~\cite{Dixon:2012yy}.

One starts by defining the \textit{Drinfeld associator} for the alphabet $X\,{=}\,\{x_0,x_1\}$ as
\begin{equation}
	Z(x_0,x_1)=\sum_{w\in X^\times}\zeta(w)w.
\end{equation}
Next, a second alphabet $Y\,{=}\,\{y_0,y_1\}$ is used, which is related to the alphabet $X$ via the monodromy equations
\begin{align}
	\label{eq:monodromy}
	y_0&=x_0,\\
	\label{eq:3.24}
	\widetilde{Z}(y_0,y_1)y_1\widetilde{Z}(y_0,y_1)^{-1}&=Z(x_0,x_1)^{-1}x_1Z(x_0,x_1).
\end{align}
In this equation, $\widetilde{ }:Y^\times\rightarrow Y^\times$ is the map that reverses the order of letters in a word, here applied to the set of words $Y^\times=\{y_0,y_1\}^\times$. The second monodromy equation (\ref{eq:3.24}) can be solved iteratively for $y_1$, where the inversions of the Drinfeld associator have to be taken as series expansions in orders of the weight of the words. The first terms of the iterative solution for $y_1$ are
\begin{align}
		\label{eq:y1}
		y_1=x_1-\zeta(3)(&2x_0x_0x_1x_1-4x_0x_1x_0x_1+2x_0x_1x_1x_1+4x_1x_0x_1x_0\notag\\ &-6x_1x_0x_1x_1-2x_1x_1x_0x_0+6x_1x_1x_0x_1-2x_1x_1x_1x_0)+\ldots\,.
\end{align}
The letter $y_1$ is only different from $x_1$ starting at weight four and a change from $x_1$ to $y_1$ involves the insertion of multiple zeta values coming from the Drinfeld associator. Using the function $\phi:Y^\times\rightarrow X^\times$ that renames $y$ to $x$, the generating series with HPLs is defined as
\begin{equation}
	L_X(z)=\sum_{w\in X^\times}H_ww,\quad \tilde{L}_Y(\overline{z})=\sum_{w\in Y^\times}\overline{H}_{\phi(w)}\widetilde{w},
\end{equation}
using the core functionals $H(w;z)=H_w$ and $H(w;\overline{z})=\overline{H}_w$. For the generating series of the svHPLs
\begin{equation}
	\CL(z)=\sum_{w\in X^\times}\CL_w(z)w
\end{equation}
the ansatz
\begin{equation}
	\CL(z)=L_X(z)\tilde{L}_Y(\overline{z})
\end{equation}
is made, yielding explicit formul\ae{} for svHPLs in terms of HPLs by comparing terms of the two series at each weight. Up to weight two, these can be found to be
\begin{subequations}
\begin{align}
		\CL_0(z)&=H_0+\overline{H}_0,\hspace{14ex} \CL_1(z)=H_1+\overline{H}_1,\\
		\CL_{00}(z)&=H_{00}+\overline{H}_{00}+H_0\overline{H}_0,\quad \CL_{01}(z)=H_{01}+\overline{H}_{10}+H_0\overline{H}_1,\\
		\CL_{10}(z)&=H_{10}+\overline{H}_{01}+H_1\overline{H}_0,\quad \CL_{11}(z)=H_{11}+\overline{H}_{11}+H_1\overline{H}_1.
\end{align}
\end{subequations}
Finally, the generating series for svHPLs $\CL(z)$ obeys two differential equations
\begin{equation}
	\label{eq:diff_eq}
	\frac{\partial}{\partial z}\CL(z)=\left(\frac{x_0}{z}+\frac{x_1}{1-z}\right)\CL(z),\quad \frac{\partial}{\partial \overline{z}}\CL(z)=\CL(z)\left(\frac{y_0}{\overline{z}}+\frac{y_1}{1-\overline{z}}\right).
\end{equation}
The connection of the svHPLs $\CL_w$ to the single-valued polylogarithms $\CG_w$, introduced in section \ref{sec:polylogszeta} and used throughout this article, can be made by the replacement $x_1\mto -x_1$. This leads to the relation
\begin{equation}
	\label{eqn:svmplrelation}
	\CG_w(z)=(-1)^{\#(x_1,w)}\CL_w(z),
\end{equation}
where $w\in X^\times$ and $\#(x_1,w)$ denotes the number of appearances of $x_1$ in $w$.


\bibliographystyle{kltmrk}
\bibliography{kltmrk}

\end{document}